\documentstyle[pra,epsf,aps]{revtex}

\begin{document}
\title{A variational description of the ground state structure \\ 
in random satisfiability problems}

\author{Giulio Biroli, R{\'e}mi Monasson, Martin Weigt\thanks{
Present address: Institut f{\"u}r Theoretische Physik, Universit{\"a}t
G{\"o}ttingen, Bunsenstr. 9, 37073 G{\"o}ttingen, Germany.}.}
\address{Laboratoire de Physique Th{\'e}orique de l'ENS,
 \thanks{Unit{\'e}  Mixte de Recherche du Centre National de
la Recherche Scientifique et de l'Ecole Normale
Sup{\'e}rieure}, 24 rue Lhomond, 75231 Paris cedex 05, France\\
biroli,monasson,weigt@physique.ens.fr}
\maketitle

\begin{abstract}
A variational approach to finite connectivity spin-glass-like models
is developed and applied to describe the structure of optimal
solutions in random satisfiability problems. Our variational scheme
accurately reproduces the known replica symmetric results and also
allows for the inclusion of replica symmetry breaking effects. For the
$3$-SAT problem, we find two transitions as the ratio $\alpha$ of
logical clauses per Boolean variables increases. At the first one
$\alpha _s \simeq 3.96$, a non-trivial organization of the solution
space in geometrically separated clusters emerges. The multiplicity of
these clusters as well as the typical distances between different
solutions are calculated. At the second threshold $\alpha_c \simeq
4.48$, satisfying assignments disappear and a finite fraction $B_0
\simeq 0.13$ of variables are overconstrained and take the same values
in all optimal (though unsatisfying) assignments. These values have to
be compared to $\alpha _c \simeq 4.27 , B_0 \simeq 0.4$ obtained from
numerical experiments on small instances.  Within the present
variational approach, the SAT-UNSAT transition naturally appears as a
mixture of a first and a second order transition.  For the mixed
$2+p$-SAT with $p<2/5$, the behavior is as expected much simpler: a
unique smooth transition from SAT to UNSAT takes place at
$\alpha_c=1/(1-p)$.
\vskip .2cm \noindent
PACS Numbers: 05.20, 64.60, 89.90 - Preprint LPTENS 99/22
\end{abstract}

\section{Introduction}

Over the last few years, the computer science community has become
increasingly aware of the occurrence of phase transitions in hard
combinatorial problems \cite{trans}. When some control parameters are
tuned, many problems of practical importance indeed exhibit drastic
changes of their behavior. The interest in such threshold phenomena
has been enhanced by the observation that instances located at phase
boundaries are the most difficult ones to solve. Even NP-complete
problems \cite{GaJo,Papa} (whose solving times are thought to grow
exponentially with their sizes) do not behave so badly far from the
threshold. As a consequence, the results of {\em worst-case}
complexity theory do not seem to be much relevant in practice and the
need for a {\em typical-case} complexity theory has clearly emerged.
Recently, the use of techniques and concepts of the statistical
physics of disordered systems combined with numerical investigations
have suggested that the nature of the transition taking place could be
related to the upsurge of complexity at the threshold \cite{nature}.
This conjecture can be best exemplified on the paradigm of
combinatorial problems showing a phase transition behavior, that is
the random $K$-Satisfiability ($K$-SAT) problem.

$K$-SAT is defined as follows.  Consider $N$ Boolean variables
$\{x_i=0,1\}_{i=1,\ldots,N}$.  Choose randomly $K$ among the $N$
possible indices $i$ and then, for each of them, choose a literal that
is the corresponding $x_i$ or its negation $\bar x_i$ with equal
probabilities one half. A clause $C$ is the logical OR of the $K$
previously chosen literals, that is $C$ will be true (or satisfied) if
and only if at least one literal is true.  Next, repeat this process
to obtain $M$ independently chosen clauses
$\{C_\ell\}_{\ell=1,\ldots,M}$ and ask for all of them to be true at
the same time (i.e. we take the logical AND of the $M$ clauses).  For
large instances ($M,N \to \infty$), numerical simulations and
mathematical analysis indicate that the probability of finding a
logical assignment of the $\{x_i\}$'s satisfying all the clauses falls
abruptly from one down to zero when $\alpha=M/N$ crosses a critical
value $\alpha_c(K)$. Above $\alpha_c(K)$, not all clauses can be
satisfied simultaneously.  This scenario is rigorously established for
2-SAT which is a polynomial problem and whose threshold $\alpha _c(2)$
equals 1 \cite{alpha2=1}. When $K\ge3$, $K$-SAT is NP-complete. Some
upper and lower bounds on $\alpha _c(K)$ have been derived and
numerical simulations have recently allowed to find estimates of
$\alpha _c$, e.g. $\alpha _c (3) \simeq 4.25-4.30$ 
\cite{Mitchell,kirk1,nature}.

When combining $p.M$ of clauses of length 3 with $(1-p).M$ clauses of
length 2, one obtains the so-called $2+p$-SAT model, which smoothly
interpolates between the instances of the easy-polynomial (2-SAT
when $p=0$) and the hard-exponential (3-SAT when $p=1$) classes
\cite{nature}.  Statistical mechanics and replica theory show that
there is a tricritical value $p_0\simeq 0.4$ separating second-order
SAT-UNSAT phase transitions for $p<p_0$ from random first-order
SAT-UNSAT phase transitions for $p>p_0$. The change of the nature of
the transition results from a change of the structure of the optimal
Boolean assignments (satisfying all clauses when $\alpha < \alpha _c
(2+p)$ or minimizing the number of violated clauses for $\alpha >
\alpha _c (2+p)$) when crossing the threshold. As shown in
\cite{nature}, the SAT-UNSAT transition results from the freezing of a
finite fraction of Boolean variables which acquire a constant value in
all optimal assignments. The emergence of such over-constrained
variables at $\alpha _c (2+p)$ appears to be continuous when $p < p_0$
and becomes strongly discontinuous above $p_0$. The existence of a
$O(N)$ backbone of over-constrained variables at the threshold above
the tricritical point has deep consequences. Indeed, a common search
algorithm such as the Davis-Putnam procedure will fail with finite
probability to correctly assert the first variable and will waste much
time in exploring empty branches of the search tree before
backtracking and correcting the early mistake. Numerical experiments
strongly support this feeling: at the threshold $\alpha _c (2+p)$, the
running time to solve an instance of the $2+p$-SAT problem behaves
polynomially with $N$ for $p \le 0.4$ and exponentially for $p \ge
0.6$ \cite{nature}.

A further understanding of the SAT-UNSAT transition undoubtedly
requires a deeper knowledge of the organization of the optimal
assignments. Information about the mutual (Hamming) distance between
solutions, the size of the backbone, etc. is indeed of high relevance
to understand and hopefully improve the efficiency of algorithms. From
a statistical physics point of view, the main difficulty stems from
the fact that $K$-SAT is naturally mapped onto a disordered spin model
with finite connectivity. Although the lack of geometrical correlation
in the clauses makes this model mean-field, the finite number of
neighbours to each spin results in much stronger local field
fluctuations and the theory is not as simple as its
infinite-connectivity counterpart.  Previous studies have shown that
even at the simplest replica symmetric (RS) level, the order parameter
describing finite-connectivity spin-glasses turns out to be a full
distribution of effective fields \cite{Beyond}.  Its determination
requires to solve a functional self-consistent equation and is far
from being easy.  The situation becomes even worse and apparently
mathematically intractable (except in some very peculiar cases
\cite{dedominicis}) when replica symmetry breaking (RSB) effects are
taken into account.

To circumvent the difficulty of solving the RS or RSB self-consistent
equations, we propose in this article a different strategy. Our claim
is that a variational approach is of high efficiency to provide very
precise results at a bearable calculation cost. Using some elementary
information about the gross physical features of the $K$-SAT model,
e.g. the existence of a backbone, we show that a RS variational
calculation is able to recover all known results and to predict new
ones (under certain assumptions) such as the value of the 
tricritical point $p_0=2/5$. In
addition, we present some new results obtained from RSB variational
calculations in both SAT and UNSAT regimes that unveil the structure
of optimal assignments in the $K$-SAT problem.  This paper is
organised as follows. In Section~II, we recall the main steps of the
statistical mechanics approach to the $K$-SAT problem
\cite{RemiRiccardo1,RemiRiccardo3}. We then explain the variational
procedure to be followed depending on the particular phase, SAT or
UNSAT, under investigation. Section~III is devoted to the analysis of
the structure of optimal configurations in the SAT phase. The
SAT-UNSAT transition is studied in Section~IV. For both Sections~III
and IV, we first focus on the replica symmetric variational solution
and then expose the additional features corresponding to replica
symmetry breaking effects. Finally, the emerging picture of the space
of solutions is summed up and some perspectives may be found in
Section~V.

\section{Statistical mechanics and variational approach}

\subsection{Replica formalism and free-energy functional}

In this section, we shall give an overview of the statistical
mechanics approach to random $K$-Satisfiability problems, see
\cite{RemiRiccardo1,Remi1} for the original works on this subject.  We
adopt the Ising-spin notion, so a true Boolean variable is mapped onto
$S_i=+1$, whereas a false variable gives $S_i=-1$.  A logical
assignment $\{S\}$ is a set of $N$ spins $S_i$ out of all $2^N$
possible configurations. We denote the (random) set of clauses by
$\{C\}$. We choose the energy-cost function ${\cal H}[\{C\},\{S\}]$ to
be the number of clauses violated by the configuration $\{S\}$
\cite{RemiRiccardo1}.  If the ground state energy is zero
(respectively strictly positive), the logical clauses are satisfiable
(resp. unsatisfiable).  The free-energy density $f$ of the resulting
spin system at a formal temperature $T$ is given by the logarithm of
the partition function
\begin{equation}
Z[\{C\}]=\sum_{\{S\}} \exp ( -{\cal H}[\{C\},\{S\}] / T )\ ,
\end{equation}
and is assumed to be self-averaging \cite{self} as the size $N$ of
the instance of the $K$-SAT problem goes to infinity. 
In order to calculate the disorder average, the replica trick
is used:
\begin{equation}
\overline{\ln Z} = \lim_{n\to 0} \partial_n \overline{Z^n}
\end{equation}
where at first a positive integer number $n$ is considered, and the
replica limit $n\to 0$ is achieved by some kind of analytical
continuation in $n$. Introducing the $2^n$ order parameters
$c(\vec{\sigma })$ equal to the fractions of ``sites'' $i$ such that
$\sigma ^{a}=S_{i}^{a}, \forall a=1,\dots ,n$ \cite{Remi1}, the
thermodynamic limit of the free-energy density can be calculated by
the saddle point method from
\begin{eqnarray}
\label{freeenergy}
f(K,\alpha ,\beta) &=& \lim_{n\to 0} \frac{1}{\beta
n}\sum_{\vec{\sigma}} c(\vec{\sigma}) \ln c(\vec{\sigma})
- \frac{\alpha}{\beta n} \ln\left[
\sum_{\vec{\sigma}_1,...,\vec{\sigma}_K} c(\vec{\sigma}_1)\cdots
c(\vec{\sigma}_K) \prod_{a=1}^n \left( 1+(e ^{-\beta}-1) \prod_{l=1}^K
\delta_{\sigma_l^a,1} \right)\right]
\end{eqnarray}
through a maximization over all normalized -- $\sum_{\vec{\sigma}}
c(\vec{\sigma}) =1$ -- and even -- $c(-\vec{\sigma}) = c(\vec{\sigma})
$ -- order parameters \cite{Beyond}. Eventually, the ground state
properties are obtained as the temperature $T=1/\beta$ is sent to zero
in (\ref{freeenergy}). Following \cite{Remi1}, the first
(respectively second) term on the r.h.s. of (\ref{freeenergy}) will be
hereafter called the effective entropy (resp. effective energy)
contribution to the free-energy.

\subsection{Simplest order parameter and replica symmetry}

Finding the saddle-point $c(\vec{\sigma })$ of (\ref{freeenergy}) is in
general a very hard task. Since the functional (\ref{freeenergy}) is
invariant under permutations of the $n$ replicas, it is possible to restrict
the variational problem to the subspace of $c(\vec{\sigma })$
with the same permutation symmetry. In this so-called replica
symmetric (RS) subspace, $c(\vec{\sigma})$ depends on
the argument $\vec{\sigma}$ only through $\sum_{a=1}^n\sigma_a$.
This allows the introduction of the generating function $P(h)$:
\begin{equation}\label{p(h)}
c(\vec{\sigma })=\int _{-\infty}^{+\infty}dh P(h)\prod _{a=1}^{n}
\left(\frac{e^{\beta h \sigma _{a}}}{e^{\beta h}+e^{-\beta h}}
\right).
\end{equation}
The normalization of $c(\vec{\sigma })$ implies the normalization of
the generating function, $\int _{-\infty}^{+\infty}dh P(h)=1$.
 Plugging this form into (\ref{freeenergy}), one can easily obtain the
analytical continuation in $n$, finally getting \cite{Remi1}
\begin{eqnarray}
\label{frs}
f_{rs}(\beta) &=&- \frac 1\beta
\int \frac{dh\ d\nu}{2\pi } e ^{i\nu h} P_{ft}(\nu)\ 
\left[1-\ln P_{ft}(\nu)\right]\ \ln(2\cosh\beta h)
\nonumber\\ &&-
\frac{\alpha}{\beta}\int \prod_{l=1}^K
dh_l\ P(h_l)\ \ln\left[ 1 + (e ^{-\beta}-1)\prod_{l=1}^K
\left(\frac{e ^{-\beta h_l}}{2\cosh \beta h_l}\right)\right]
\end{eqnarray}
where $P_{ft}(\nu)=\int dh\ e ^{-i\nu h}P(h)$
denotes the Fourier transform of the generating function $P(h)$.
The free energy (\ref{frs}) now has to be optimized with respect to 
$P(h)$.

To understand the physics hidden in this approach, it is useful to
consider the Boolean magnetizations $m_{i}=\ll  S_{i} \gg  $, where $\ll 
. \gg  $ denotes the Gibbs average with Hamiltonian ${\cal H}$ at fixed
disorder $\{ C\}$, see Section II.A .  An effective field $h_{i}$ is 
associated to each local magnetization $m_{i}$ through the relation
$m_{i}=\tanh (\beta h_{i})$. Within the RS framework, the order
parameter $P(h)$ is simply the histogram of the effective fields,
\begin{equation}
P(h)=\frac{1}{N}\sum_{i=1}^{N}\overline{\delta (h-h_{i})  }
\qquad ,
\end{equation}
where the overbar denotes the average over the random choices of clauses
$\{C\}$. At very low temperature, effective fields are related to elementary
excitations around ground state configurations.

\subsection{More sophisticated order parameters and replica symmetry breaking}

A corollary of Ansatz (\ref{p(h)}) is that the Hamming distance $d$ between
any two assignments ({\it i.e.} the number of variables which are
different in the two configurations) weighted with the Gibbs measure 
almost surely equals 
\begin{equation}
d_{rs} = \frac 12 - \frac 12 \int dh \; P(h) \; \left( \tanh \beta h 
\right)^2 \qquad ,
\label{drs}
\end{equation}
once divided by $N$. In other words, on the $N$-dimensional hypercube
whose vertices are the Boolean configurations, all relevant
assignments belong to a single {\em cluster} of typical diameter $d
_{rs}\cdot N$. On general grounds, there is no {\em a priori} reason
to trust this simple picture. At zero temperature for instance, there
could well exist a non trivial geometrical organization of the space
of solutions to the SAT or MAX-SAT problem which would give rise to a
non-trivial ({\em i.e.} not fully concentrated) probability
distribution for $d$. The simplest and immediate extension of
(\ref{drs}) corresponds to a bimodal distribution for $d$, with two
peaks in $d_0$ and $d_1 (< d_0)$. The corresponding picture on the
hypercube of configurations is that solutions are now gathered into
different clusters having average internal diameter $d_1\cdot N$ and
being separated by a typical distance $d_0\cdot N$.

Let us label these clusters, also called pure states \cite{Beyond},
by a new index $\Gamma$. In a given cluster $\Gamma$, the magnetizations
$m_{i}^{\Gamma }$ can be calculated as the average values of the spins $S_i$ 
over the ${\cal N}_\Gamma$
assignments belonging to $\Gamma$. As before, it is convenient
to consider the effective fields $h_{i}^{\Gamma }$ through the relations
$m_{i}^{\Gamma}=\tanh (\beta h_{i}^{\Gamma })$. These effective
fields fluctuate 
\begin{itemize}
\item from  state to state:
For a given site $i$, the effective fields $h_{i}^{\Gamma}$ depend on
the cluster $\Gamma$. We introduce the histogram $\rho_i
(h)=\sum_{\Gamma}{\cal N} _\Gamma\;\delta(h-h_i
^{\Gamma})/\sum_{\Gamma}{\cal N} _\Gamma$ to take these fluctuations
into account.
\item from spin to spin:
In turn, $\rho_i (h)$ explicitely depends upon the index $i$ of the
variable it is related to. This multiplicity of field histograms is 
encoded in a functional probability distribution ${\cal P}[\rho]
=\sum_{i=1}^N \delta[\rho(h)-\rho_i (h)]/N$ over
the set of possible $\rho (h)$.
\end{itemize}

Within the replica formalism exposed in Section II.A, the above picture
corresponds to the first step of Parisi's hierarchical replica
symmetry breaking (RSB) scheme \cite{Beyond}. 
The RSB order parameter $c(\vec{\sigma })$ reads \cite{Remi1}
\begin{equation}\label{crsb}
c(\vec{\sigma })=\int {\cal D}\rho{\cal P}[\rho ]
\prod _{b=1}^{n/m} \left\{ 
\int_{-\infty }^{+\infty }dh \rho (h) \prod _{a=1+(b-1)m} ^{b\; m}
\left( \frac{e^{\beta h \sigma _a}}{e^{\beta h}+e^{-\beta h}} \right)
\right\} \quad . 
\end{equation}
With the above Ansatz, the analytical continuation $n\to 0$ can be performed 
and the resulting free-energy (\ref{freeenergy}) has to be 
optimized over $P[\rho (h)]$, see \cite{Remi1}. The parameter $m$ in
(\ref{crsb}) determines the relative importances of $d_0$ and $d_1$ 
\cite{Remi1,Beyond}.

\subsection{Variational approach}

The direct way to complete the calculation of free energy
(\ref{freeenergy}) within the replica symmetric or the one-step broken
approximation would be the following: a variation of
(\ref{freeenergy}) with respect to the order parameters yields
functional equations for $P(h)$ or ${\cal P}[\rho(h)]$. In the replica
symmetric case, this equation could be solved in \cite{RemiRiccardo3}
by a class of distributions consisting of a larger and larger number
of Dirac peaks. In the replica symmetry broken case, only the very
simplest possible solution could be obtained in \cite{Remi1}. The 
evaluation of any more involved solution seems a hopeless task due
to the complexity of the saddle point equations \cite{pspindedom}.

In this paper a different route will be chosen \cite{varia9}. Based on
physical grounds, some simple trial functions for $P(h)$ or ${\cal
P}[\rho(h)]$ will be proposed. These functions only depend on a small
number of parameters; this fact significantly reduces the
complexity of the problem. In the replica symmetric case, the exact
results of \cite{RemiRiccardo3} can be reproduced within a precision
of less then one percent. In the replica symmetry broken case, new
results can be obtained which are far beyond the solution given in
\cite{Remi1}.

\subsection{Zero temperature limit and scaling of the effective fields}

The phase transition in $K$-SAT separates a low $\alpha $ regime in
which all variables are typically under-constrained (SAT regime) from a
high $\alpha $ regime in which a finite fraction of variables is
typically over-constrained (UNSAT regime).

Variables can be under-constrained when they do not appear in any
clauses, or more generally when the minimal number of violated clauses
is independent of their possible assignments (true or false). In the
language of statistical physics, such under-constrained variables
correspond at low temperature $T$ to spins $S_i$ submitted to
effective fields $h_i$ vanishing linearly with $T$: $h_i = T
. z_i$. This way, their magnetizations $m_i = \tanh (\beta h_i ) =
\tanh z_i$ are different from $\pm 1$ in the ground state.  These
unfrozen spins do not contribute to the energy when $T\to 0$, but only
to the entropy. In the SAT phase, effective fields are expected to
show this behavior for low temperatures.

Conversely, over-constrained variable correspond to spins $S_{i}$
seeing effective fields $h_i$ that remain finite, {\em i.e.} of the
order of one in the zero temperature limit. The excitation energy to
flip any of these spins $S_{i}$ is finite and the spins are frozen in
up or down directions depending on the signs of the associated fields
$h_i$.  In the zero temperature limit the only contribution to the
energy comes from these frozen spins. To study the SAT-UNSAT
transition, one therefore has to focus onto the probability
distribution of effective fields on the scale of $O(1)$. Note that on
this scale, the effective fields corresponding to unfrozen spins
vanish and give rise to a Dirac peak centered at zero; the weight of
this $h=0$ peak is precisely the fraction of under-constrained spins.


\section{The satisfiable phase}

\subsection{Replica symmetric approximation}

As already discussed in the previous section, the interesting quantity
to be calculated in the satisfiable phase ($\alpha < \alpha _c$) is
the ground state entropy density $s=-\lim_{\beta\to\infty}\beta f$ in
the satisfiable phase ($\alpha < \alpha _c$). In the replica symmetric
approximation, the entropy $s$ reads, according to equation
(\ref{frs}),
\begin{eqnarray}
\label{entropie}
s_{rs} &=& \alpha \int \prod_{l=1}^K dz_l \tilde{P}(z_l) \ln\left[ 1 -
\prod_{l=1}^K \frac{e ^{-z_l}}{2\cosh z_l} \right]
 +\int dz\  \tilde{P}(z) \ln[ 2\cosh z] \nonumber\\
&& - \int \frac{dz\ d\nu}{2\pi} e ^{iz\nu} \tilde{P}_{ft}(\nu) 
\ln[\tilde{P}_{ft}(\nu)] \ln[ 2\cosh z]\ ,
\end{eqnarray}
where $z=\beta h$ is the rescaled effective field of order one, see
Section II.E. As a consequence, the distribution $\tilde{P}(z)$ has
a finite mean and variance in the limit $\beta\to 0$.  As in the
previous Section, $\tilde{P}_{ft}(\nu )$ denotes the Fourier transform
of $\tilde{P}$.

\begin{figure}[bt]
\centerline{    \epsfysize=9cm
        \epsffile{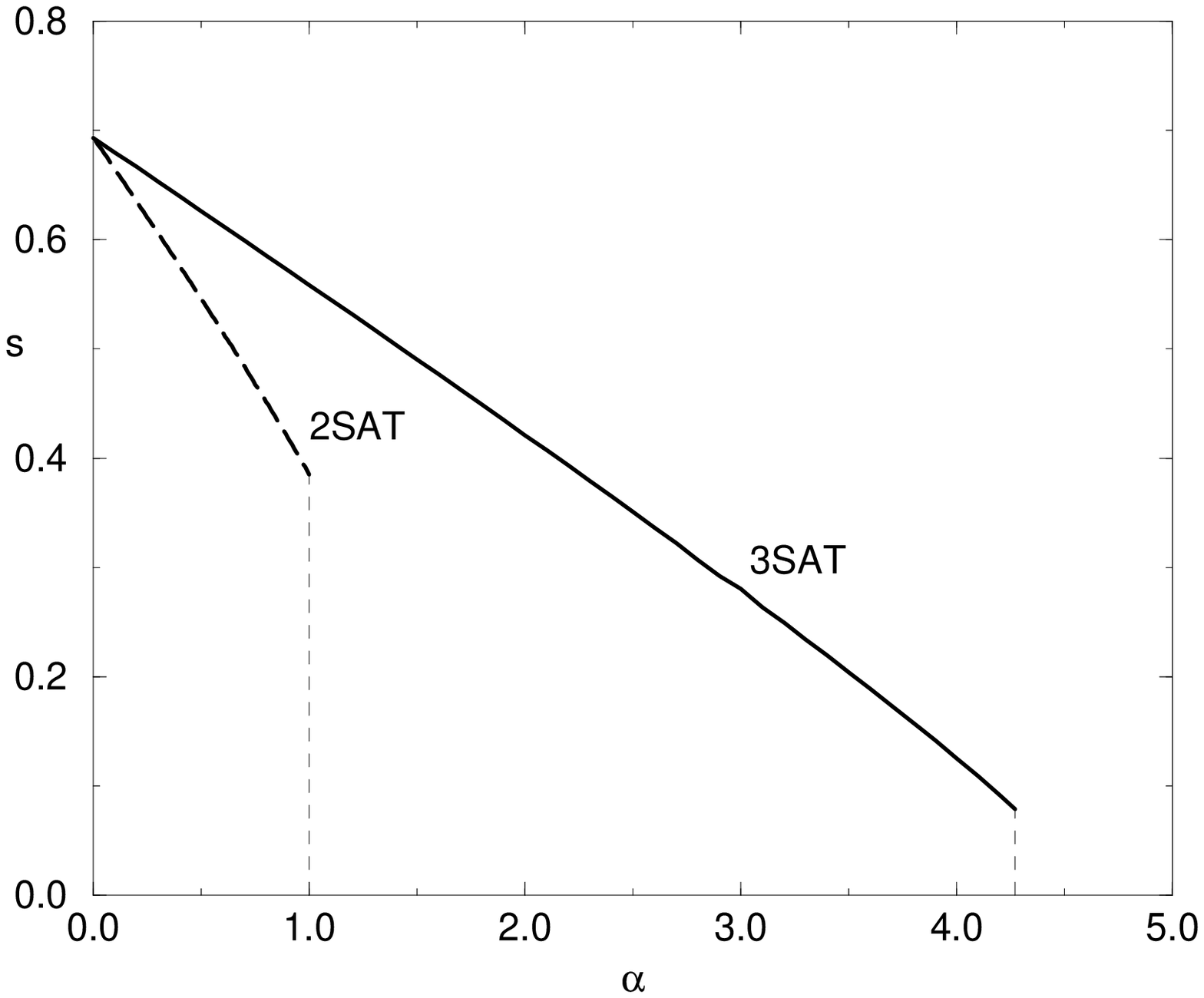}}
\caption{Variational entropy $s$ of the solutions for the 2-SAT (bold dashed 
line) and the 3-SAT (full line) problems as functions of the ratio
of clauses per variable $\alpha$. The curves are practically
indistinguishable from the full RS results in \protect\cite{RemiRiccardo1}. 
The vertical dashed lines
indicate the threshold $\alpha _c (2)=1$ and $\alpha _c (3) \simeq
4.27$ \protect\cite{nature}.}
\label{fig:entropy}
\end{figure}

\begin{figure}[bt]
\centerline{    \epsfysize=9cm
        \epsffile{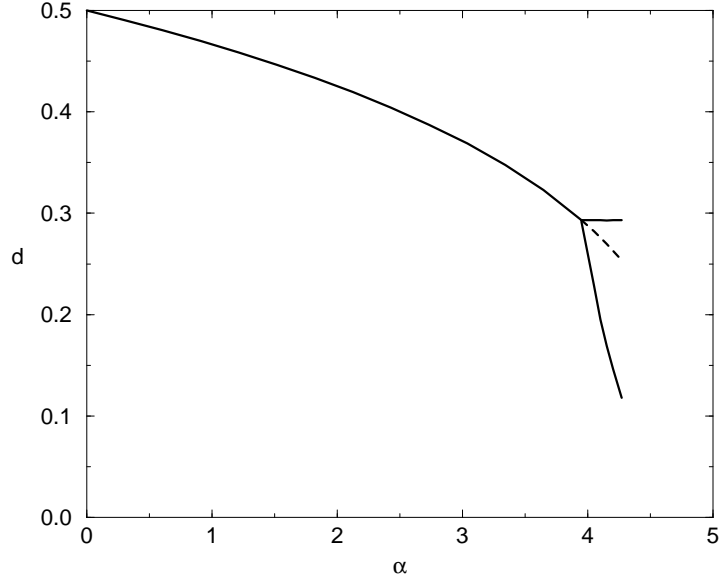}}
\caption{Typical Hamming distance between two solutions of a random
3-SAT problem. Whereas there is only one distance $d$ in the replica
symmetric phase $0\leq\alpha<\alpha_s \simeq 3.96$ which is monotonously
decreasing with the number of clauses per variables $\alpha$, we find
two characteristic 
lengths in the replica symmetry broken case $\alpha _s < \alpha <
\alpha_c \simeq 4.25-4.30$ \protect\cite{nature,Mitchell}.  
The Hamming distance $d_0$
between two clusters of solutions (upper line) remains almost constant 
with $\alpha$, whereas the entropy loss is mostly due to a shrinking
of the typical size $d_1$ of the clusters themselves (lower curve). 
The dashed line
denotes the continuation of the replica symmetric result.}
\label{fig:distance}
\end{figure}

We start with a simple Gaussian Ansatz for the rescaled field distribution,
\begin{equation}
\label{Ansatzrs}
\tilde{P}(z) =  G_\Delta(z),
\end{equation}
where $G_\Delta(z)$ denotes a Gaussian distribution with zero mean and
variance $\Delta$.  Note that $\tilde{P}(z)$ is expected to be even
due to the symmetry of the disorder distribution.  In the case of
infinite connectivity spin glasses eq. (\ref{Ansatzrs}) would give the
exact distribution, cf. \cite{Beyond}, but due to the finite
connectivities in $K$-SAT effective fields are not necessarily
Gaussianly distributed. We find the expression
\begin{eqnarray}
\label{entropie2}
s_{rs} &=& \alpha \int \prod_{l=1}^K Dz_l \ln\left[ 1 -
\prod_{l=1}^K \frac{e ^{-\sqrt{\Delta} z_l}}{2\cosh \sqrt{\Delta}z_l} 
\right] + \int Dz \left( \frac{3-z^2}{2} 
\right) \ln[ 2\cosh\sqrt{\Delta} z ]\ ,
\end{eqnarray}
which has to be optimized numerically with respect to the variational
parameter $\Delta$. Hereafter, $Dz = G_1(z) \; dz$ denotes the
Gaussian measure with zero mean and variance one.  For $\alpha=0$ the
variational parameter is found to be $\Delta=0$, and the entropy
follows to be $s_{rs}=\ln 2$: there are no clauses and the solution
space coincides with the full phase space of the model. For increasing
$\alpha$, the entropy diminishes due to the growing number of
constraints, see figure \ref{fig:entropy}.  Our results are
practically indistinguishable from the exact expansion of $s_{rs}$ in
powers of $\alpha$ performed in \cite{RemiRiccardo1}.  In
fig.\ref{fig:distance}, we show the typical Hamming distance $d_{rs}$
(\ref{drs}) between two solutions. $d_{rs}$ monotonously decreases
from $d=0.5$ at $\alpha=0$. This behavior signals a concentration of
the solutions in configuration space.


In order to test the robustness of the results obtained with help of
Ansatz (\ref{Ansatzrs}) we have repeated the above calculation with an
exponential Ansatz for $\tilde{P}(z)$.  The values of the entropy were
changed by less than $1\%$. We have also taken into account the
presence of free spins through the Ansatz
\begin{equation}
\label{po9}
\tilde{P}(z) = (1-A)\; \delta(z) + A\; G_{\Delta} (z) 
\end{equation}
for the rescaled distribution. The Dirac peak accounts for the
variables which are not contained in the clauses as well as for spins
having an effective field going to zero faster than linearly with the
formal temperature $T$.  Therefore $1- \exp( -\alpha K )$, {\it i.e.}
the fraction of variables not appearing in any clause, constitutes a
rigorous upper bound for $A$ which was explicitely violated by the
Ans{\"a}tze considered so far.  For 2-SAT at $\alpha=1$, we find
$A=0.57 \pm 0.01$ (and $0.71 \pm 0.02$ if the Gaussian in (\ref{po9})
is again replaced by an exponential distribution), which has to be
compared to the bound $1- \exp( -2)=0.865$. The strong dependence of
$A$ on the non-zero field part of the Ansatz probably results from the
inclusion of small but non-zero rescaled fields into the Dirac
peak. For 3-SAT, differences are less drastic: whereas the upper bound
is almost 1, the above Ansatz gives $A=0.94 \pm 0.02$ (and an $A$
numerically indistinguishable from 1 for the exponential distribution)
at $\alpha=4$.  In all cases, all these variations affect the entropy
value by $1\%$ at the most.

\subsection{The replica symmetry breaking transition}

We have already underlined in section II.C that the replica symmetric
Ansatz is unable to reflect any non-trivial organization of
the optimal assignments space.  To investigate the ground state
structure of $K$-SAT, we thus consider a one-step replica
symmetry broken (RSB) Ansatz. According to the discussion of Section II.C,
we choose
\begin{equation}
\label{AnsatzrsB}
{\cal P}[\rho ( z ) ] = \int_{-\infty}^\infty dz\  G_{\Delta_0 }(z) 
\ \delta\left[ \rho(\tilde{z}) - \frac{ G_{\Delta_1 }(\tilde{z}-z)
(2\cosh \tilde{z})^m}{
\int dz'\ G_{\Delta_1 }(z'-z) (2 \cosh z')^m }
\right]\ ,
\end{equation}
which coincides with the exact one-step expression for
infinite-connectivity spin glass models. As in the RS hypothesis
(\ref{entropie}), $z=\beta h$ is a rescaled field which remains of
order one in the zero-temperature limit. The detailed calculation of
the variational RSB ground state entropy $s_{rsb}$ in the satisfiable
phase is exposed in appendix A. The result reads
\begin{eqnarray}
\label{entropyrsB}
s_{rsb} &=& \frac{\alpha}{m} \int \prod_{l=1}^K Dz_l\ \ln
\frac{\int \prod_l D\tilde{z}_l \left(\prod_l 2\cosh(\sqrt{\Delta_1}
\tilde{z}_l + \sqrt{\Delta_0} z_l) - \prod_l \exp\{\sqrt{\Delta_1}
\tilde{z}_l + \sqrt{\Delta_0} z_l\}\right)^m}{\int \prod_l
D\tilde{z}_l \left(\prod_l 2\cosh(\sqrt{\Delta_1}
\tilde{z}_l + \sqrt{\Delta_0} z_l)\right)^m}\nonumber\\
&& + \frac{1}{m} \int Dz \ \ln \int D\tilde{z}
\left(2\cosh(\sqrt{\Delta_1}\tilde{z}+ \sqrt{\Delta_0} z)\right)^m
\nonumber\\
&& - \int Dz \frac{ \int D\tilde{z} 
\left(2\cosh(\sqrt{\Delta_1}\tilde{z}+ \sqrt{\Delta_0} z)\right)^{m-1}
\sinh(\sqrt{\Delta_1}\tilde{z}+ \sqrt{\Delta_0} z) 
\ (\sqrt{\Delta_1}\tilde{z}+ \sqrt{\Delta_0} z) }{\int D\tilde{z} 
\left(2\cosh(\sqrt{\Delta_1}\tilde{z}+ \sqrt{\Delta_0} z)\right)^m}\ .
\end{eqnarray}
This quantity has to be optimized with respect to the
variational parameters $\Delta_{0}, \Delta _1$ and $m$. The numerical problem
in calculating the solutions of the three equations
\begin{equation}
\label{saddlersB}
0=\frac{\partial s_{rsb}}{\partial \Delta_0}
=\frac{\partial s_{rsb}}{\partial \Delta_1}
=\frac{\partial s_{rsb}}{\partial m}
\end{equation}
for 3-SAT consists in the sixfold integration in the first term
of (\ref{entropyrsB}). It is much easier to determine the critical $\alpha_s$
where the first nontrivial solution of these equations can be found.
In principle, due to the continuity of the entropy $s_{rsb}$ at this
transition, there are two possible scenarios\cite{Ga},
\begin{itemize}
\item A continuous transition in $\Delta_0\to\Delta_{rs}$,
$\Delta_1\to 0$ takes place, where $\Delta_{rs}$ here and in
the following denotes the replica symmetric value. This could be 
connected to a nontrivial $m_s$ with $0<m_s<1$. 
\item A jump in $\Delta_1$ towards a nontrivial value $>0$ takes place
at the transition. To guarantee the continuity of $s_{rsb}$, such
a transition has to happen at $m=1$ and $\Delta_0=\Delta_{rs}$.
\end{itemize}
In the following two subsections we will consider both possibilities.
Whereas a transition of the first type can be found at a certain
$\alpha_s$, the second possibility can be ruled out. The value of
$\alpha_s$ found this way constitutes an upper bound for the exact
threshold.  We can indeed not exclude that, by taking into account a
larger variety of density functionals (\ref{AnsatzrsB}), a non-trivial
RSB solution could appear already at smaller values of $\alpha$.

\subsubsection{The continuous transition}

In order to determine the critical value $\alpha_s$ for the replica
symmetry breaking transition inside the SAT phase, we have to
explicitly use equation (\ref{saddlersB}),
\begin{equation}
\label{saddle12rsB}
\frac{\partial s_{rsb}}{\partial \Delta_1} =0 \ ,
\end{equation}
and expand it to first order in $\Delta_1$. As a result of the
expansion, the interior integrals over the $\tilde{z}_l$ in
(\ref{entropyrsB}) can be carried out analytically, leaving only three
integral to be evaluated numerically. At the zeroth order,
the replica symmetric saddle point equation for $\Delta_0$
is recovered. At the first order,
the coefficient of the linear term in $\Delta _1$ vanishes at
\begin{equation}
\alpha_s=3.955\pm  0.005,
\end{equation}
thus allowing a non-zero solution for $\Delta_1$ to develop. This
value is in surprisingly good agreement with a critical slowing down
found numerically by Svenson and Nordahl \cite{Sv}. They considered a
simple zero-temperature Glauber dynamics for random satisfiability and
coloring problems. In the case of 3-SAT this dynamics showed an
exponential relaxation down to (almost) zero energy density for
$\alpha<4$, whereas the relaxation became algebraic for 
$\alpha>4$ -- converging towards non-zero energy. As we shall see in
Section III.C, the increase of $\Delta_1$ at $\alpha _s$ coincides
with the emergence of a non trivial structure of the optimal
assigments of a typical 3-SAT instance. Due to the continuous nature
of the transition at $\alpha_s$ it is probable that higher-lying,
metastable states blow up simultaneously with the breaking up of the
ground state structure.  If it were so, the relationship between
Svenson's and Nordahl's result and the static transition at $\alpha_s$
could perhaps be explained along the lines developed in the context of
the off-equilibrium dynamics of spin-glasses\cite{cuku}.

In order to calculate also  $m_s$ at the transition, we
have to take into account either the second order terms of equations
(\ref{saddle12rsB}) slightly above $\alpha_s$, or to explicitely solve
the full saddle point equations (\ref{saddlersB}) in the limit
$\alpha\to\alpha_s ^+$. We have followed the second route
\cite{numerics} and found $m_s\approx 0.8$.  The corrections to the
entropy of solutions is however very weak, $s_{rs}=0.917$ while
$s_{rsb}=0.911$ at $\alpha=4.2$.

For 2-SAT, no such transition can be found before the
SAT-UNSAT-transition. A numerical investigation of the $2+p$-SAT model
makes us conjecture that the existence of a replica symmetry breaking
transition within the SAT phase is related to the appearance of a
discontinuous SAT-UNSAT transition, see Section~IV.

\subsubsection{Nonexistence of a discontinuous transition}

As already discussed above, one could also imagine a discontinuous
transition with a jump in $\Delta_1$ even at a lower value of
$\alpha$. Due to the continuity of the ground-state entropy, we would
expect this transition to be continuous in $m$, i.e. to happen at 
$m=1$.

\begin{figure}[hbt]
 \centerline{   \epsfysize=9cm\epsffile{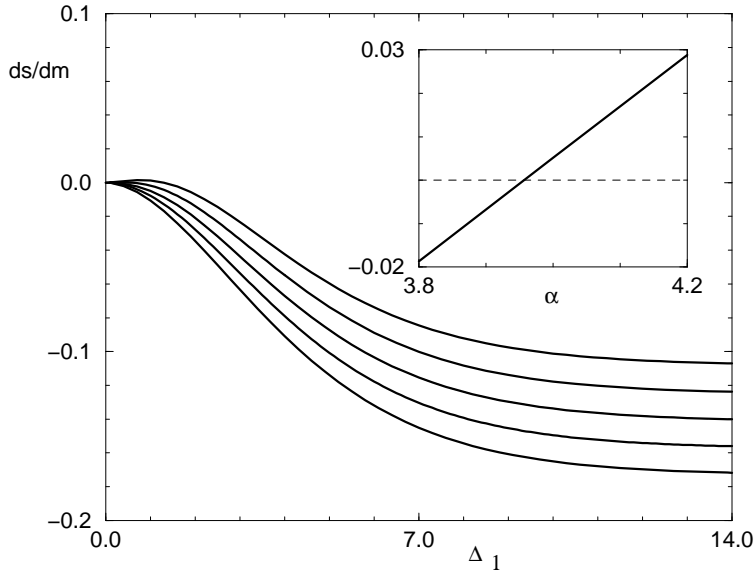} 
        }
\caption{$\partial_m s_{rsb}(m=1,\Delta_0=\Delta_{rs},\Delta_1)$ 
as function of $\Delta_1$. The figure shows the curves for
$\alpha=3.7,3.8,3.9,4.0,4.1$ (from bottom to top), illustrating the
non-existence of a discontinuous transition. Inset: the 
second derivative $\partial _{\Delta _1}^2( 
\partial_m s_{rsb})$ vanishes at $\alpha_s=3.955\pm  0.005$.}
\label{fig:dfdm}
\end{figure}

The most interesting saddle point equation to be considered here is
hence the $m$-equation. Exactly at the transition, a non-zero solution
for $\Delta_1$ should show up in the equation
\begin{equation}
\label{msaddle}
0=\frac{\partial s_{rsb}}{\partial m}
(m=1,\Delta_0=\Delta_{rs},\Delta_1)\ .
\end{equation}
The other two variational equations (\ref{saddle12rsB}) are
automatically fulfilled at this point.  In figure \ref{fig:dfdm}, we
display $\partial_m s_{rsb}(m=1,\Delta_0=\Delta_{rs},\Delta_1)$ for
several values of $\alpha$ as a function of $\Delta_1$.  For
$\alpha<\alpha_s$, i.e. below the continuous transition found in the
last subsection, the function is monotonously decreasing with
$\Delta_1$, towards some finite asymptotic value. This clearly rules
out any discontinuous transition in $\Delta_1$ -- the only zero of
this function lies at $\Delta_1=0$.  At the continuous transition the
behavior in the vicinity of $\Delta_1=0$ changes. The sign of the
second derivative of $\partial_m s_{rsb}$ with respect to $\Delta_1$
changes whereas the first derivative is always zero. This confirms
again the local instability of the replica symmetric solution leading
to the continuous transition found in the previous paragraph.

\subsection{Multiplicity of clusters}

In this section, we give a geometrical interpretation of the RSB 
transition.  Above $\alpha_s \simeq 3.96$, the
ground state configurations are divided into an
exponential number of well separated clusters.  We are interested in the
distribution of the entropy densities $s$ of these clusters, {\em
i.e.} we want to count the clusters containing $\sim e^{N s}$
satisfying assignments.  This number will be denoted by $e^{N \omega
(s)}$. The quantity $\omega(s)$ is hereafter referred to as the {\it
multiplicity} of $s$. By definition of $\omega$,
\begin{eqnarray}
\label{moments}
Z_m &:=& \int ds\; e^{N \omega(s)} \left( e ^{N s} \right) ^m \\
&=& \lim _{\beta \to \infty} 
\sum_{\Gamma} \left( \sum_{\{S_i\}\in {\Gamma}}
    e ^{ -\beta {\cal H}[\{C\},\{S\}]} \right)^m \label{mom2}\ ,
\end{eqnarray}
where $\Gamma$ denotes the clusters and ${\cal H}$ the energy cost 
function of Section II.A. $m$ is now a control parameter that can be
varied to obtain $\omega(s)$. A straightforward 
calculation of (\ref{moments}) show that $\tau(m) = \log Z_m /N$  
is simply the Legendre transform of $\omega (s)$ in the large $N$
limit, 
\begin{equation}
\label{complexity}
 \tau(m) = \lim _{N\to \infty} \frac 1N \log Z_m =
{\mbox{extr}}_{s} [ \omega(s) + m s ] \qquad .
\end{equation}
Thus to access the multiplicity $\omega $, we resort to the
calculation of $\tau$, following closely the lines of \cite{Mo2} (see
also \cite{FrPa} and \cite{MoKa} for related calculations on the
$p-$spin glass model and neural networks). We use again the replica
trick, $\overline{\ln Z_m} = \lim_{n\to 0} \partial_n
\overline{(Z_m)^n}$, and represent also the $m$-th power in
(\ref{mom2}) by a positive integer-valued $m$. This leads to $n.m$
replicas of the original system obeying the one-step RSB
algebra\cite{Mo2}. Within our Gaussian variational scheme, we easily
find
\begin{equation}
\tau(m) = m\  {\mbox{extr}}_{\Delta_0,\Delta_1}\ 
[s_{rsb}(m,\Delta_0,\Delta_1)]
\end{equation}
where $s_{rsb}$ is given by (\ref{entropyrsB}).  Consequently, the
dominant clusters considered so far and obtained by optimization of
$s_{rsb}$ over $m$ have zero multiplicity $\omega = -m^2 \partial
s_{rsb}/\partial m=0$, {\em i.e.}  their number is less than
exponential in the system size $N$.  Simultaneously, there exist
exponentially numerous clusters with lower entropies such that the
total number of satisfying assigments they contain remains much smaller
than $e^{N s_{rsb}}$.  These subdominant states appear as soon as
$\alpha$ gets larger than $\alpha_s$. Below this transition there is
no positive multiplicity at all; almost all solutions are collected in
one large cluster.

\begin{figure}[bt]
\centerline{    \epsfysize=9cm
        \epsffile{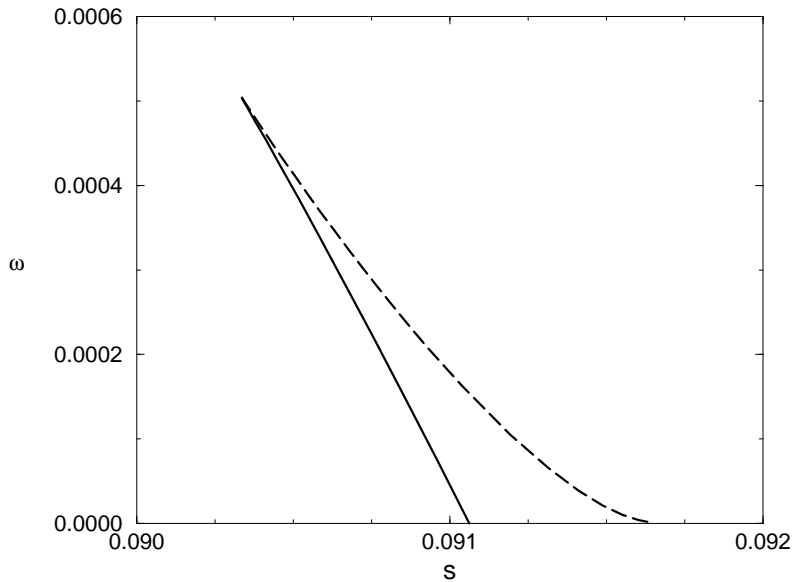}}
\caption{Multiplicity $\omega(s)$ of states with entropy $s$ at
$\alpha=4.2$. The curve intersects the zero multiplicity axis at
$s_{rsb} \simeq 0.911$ and $s_{rs} \simeq 0.917$.
The full line shows the reliable part of the curve. Along the
dashed line, the second derivative of $s_{rsb}$ is positive and the
one-step replica symmetry broken Ansatz is not longer valid. The
curves for other values of $\alpha> \alpha _s$ are qualitatively similar.}
\label{fig:compl}
\end{figure}

In figure \ref{fig:compl} we show the results for $\alpha=4.2$ (we
have checked the qualitative similarity of the curves for other values
of $\alpha$) obtained through a numerical solution of the variational
equations in $\Delta_0$ and $\Delta_1$. At a certain $s = s_{rsb}
\simeq 0.911$ the multiplicity becomes positive, and the curve starts
with slope $-m_{rsb}$, where $m_{rsb} \simeq 0.72$ is the value of $m$
that optimizes $s_{rsb}$.  At $m=0$, {\it i.e.} where the slope of
$\omega$ over $s$ vanishes, we find again the replica symmetric
entropy $s_{rs} \simeq 0.917$ calculated at the beginning of this
Section. However, the curve is only reliable up to the cusp: There,
the second derivative $d^2 s_{rsb}/dm^2$ changes sign and the
corresponding variational solution becomes unstable \cite{Ni}. Note
also that the RS entropy $s_{rs} = \tau (1)$ is found back at $m=1$ in
the unphysical negative $\omega$ region. As happens also in the
case of the spherical $p-$spin glass, there could already be an
instability of the one-step solution due to instable replicon modes
\cite{DoKoTe}, but the deviations from the given curve are expected to
be very weak.

Although the order of magnitude of the multiplicity calculated above is
small, some drastic changes take place at $\alpha _s$. This can best
seen on the  typical Hamming distances between solutions. 
The quantity
\begin{equation}
d_1= \frac{1}{2}-\frac{1}{2}\int Dz\  \frac{\int D\tilde{z}
\cosh^m(\sqrt{\Delta_0}z+\sqrt{\Delta_1}\tilde{z})
\tanh^2(\sqrt{\Delta_0}z+\sqrt{\Delta_1}\tilde{z})}{\int D\tilde{z}
\cosh^m(\sqrt{\Delta_0}z+\sqrt{\Delta_1}\tilde{z})}
\end{equation}
describes the average distance between two solutions inside
the same cluster, whereas 
\begin{equation}
d_0= \frac{1}{2}-\frac{1}{2}\int Dz\ 
\left[ \frac{\int D\tilde{z}
\cosh^m(\sqrt{\Delta_0}z+\sqrt{\Delta_1}\tilde{z})
\tanh(\sqrt{\Delta_0}z+\sqrt{\Delta_1}\tilde{z})}{\int D\tilde{z}
\cosh^m(\sqrt{\Delta_0}z+\sqrt{\Delta_1}\tilde{z})} \right]^2
\end{equation}
stands for the distance between two clusters. The results for the
thermodynamically dominant states are shown in figure
\ref{fig:distance}. We observe that $d_0$ is almost constant in
$\alpha$, {\it i.e.} the relative positions of the clusters in
configuration space remain roughly unchanged if new clauses are added
to a given sample. In contrast to this behavior, the disappearance of
solutions as $\alpha$ grows is accompanied with  a rapid
decrease of the cluster diameter $d_1$.

\subsection{Breakdown of the scaling of the effective fields}

All the Ans{\"a}tze we have used in the description of the SAT phase
were based on the assumption of effective fields linearly vanishing in
the zero-temperature limit. As we have argued, this scaling is no
longer valid above the SAT-UNSAT transition. So we can extract a first
variational estimate of the critical value $\alpha_c$ from the
divergence of $\Delta$ in the replica symmetric and of $\Delta_1$ in
the replica symmetry broken case. Compared with the numerical value
$\alpha_c=4.25-4.3$ \cite{Mitchell,kirk1,nature} the resulting values
4.76 (RS), resp.  4.66 (RSB) are rather crude approximation. The
replica symmetric value can already be improved by taking (\ref{po9})
instead of (\ref{Ansatzrs}). We find $\alpha_c=4.622$ with $A=0.935$,
which is rather similar to the iterative replica symmetric result in
\cite{RemiRiccardo1}. As we shall show in the next Section, physically
more elaborate approximations are needed to obtained better results
for $\alpha_c$.


\section{The SAT-UNSAT transition}

\subsection{Replica symmetric calculation}\label{rs}

\subsubsection{Variational RS free-energy}\label{RSVP}

According to Sections II.B and II.E, we propose the following variational
(replica symmetric) field distribution in the zero temperature limit 
\begin{equation}\label{prsvar}
P(h)=(1-B)\delta (h)+\frac{B}{\sqrt \Delta} \; \Phi \left(
\frac{h}{\sqrt \Delta}\right).
\end{equation}
$\Phi(x)$ is an even and decreasing probability distribution with 
argument $x=O(1)$. $B$ denotes the fraction of frozen spins and $\Delta$
($=O(1)$ when $T\to 0$) the typical squared magnitude of the effective
fields acting on frozen spins.

Once $\Phi (x)$ has been chosen, we plug the trial variational
function (\ref{prsvar}) into (\ref{frs}). The resulting variational
problem involves two parameters $B$ and $\Delta$ only and is
therefore considerably simpler than the initial one. In the zero
temperature limit we obtain from (\ref{frs}),
\begin{eqnarray}\label{F2+p1}
f_{rs}(B, \Delta ,\alpha ,p ) &=& -2\; \sqrt \Delta \left(\frac{B}{\pi} \int
_{0 }^{+\infty} \frac{d\nu}{\nu } \; \Phi _{ft} '(\nu ) \ln
[1-B+B \;\Phi _{ft} (\nu) ]- \right.
\nonumber \\ && \left. \alpha \;
\int_{0}^{1/{(2 \sqrt \Delta) }}dh \left\{ (1-p) \;B^2\; \left[ \Phi _{cc} (h)
\right] ^2 + p\; B^3\; \left[ \Phi _{cc} (h) \right] ^3 \right\} \right)
\end{eqnarray}
where
\begin{eqnarray}
 {\Phi} _{ft} (\nu )&=&\int_{-\infty}^{+\infty}dx\; e^{-ix\nu
} \; \Phi(x) \label{qw2} \\
\Phi _{cc} (h) &=&\int_{h}^{+\infty}dx \; \Phi (x) \label{hphi}
\end{eqnarray}
are respectively the Fourier transform and the complementary
cumulative function of $\Phi$. The above free-energy (\ref{F2+p1})
corresponds to the 2+p-SAT problem which smoothly
interpolates between 2-SAT ($p=0$) and 3-SAT ($p=1$), cf. section I. 
For the sake of completeness, we give in Appendix~B the derivation of 
(\ref{F2+p1}) for the special case of a Gaussian distribution $\Phi
(x) = G_1 (x)$. 

As we shall show below, within this simplified version of the
variational problem it is possible to obtain results which are only
slightly different from the ones obtained from the best replica
symmetric solution \cite{RemiRiccardo1}.  The simplicity of this
approach leads to a more transparent description of the SAT-UNSAT
transition.


\subsubsection{A smooth transition: the 2-SAT problem}\label{2satrs}

We start with $p=0$,{\em i.e.} the 2-SAT case. From previous numerical and
analytical studies, it is known that the fraction of frozen spins is
continuous at the transition and is zero for $\alpha < \alpha _{c}$
\cite{nature,RemiRiccardo1}.  Actually a numerical analysis of $f_{rs}
(B,\Delta,\alpha,0)$ excludes the possibility of a first order
transition in $B$ and $\Delta$. To locate the critical value of
$\alpha$, we expand $f_{rs} (B, \Delta,\alpha,0)$ around $B=0$ and
$\Delta=0$.  To the leading order and neglecting irrelevant terms in
$\Delta$, we find
\begin{equation}
f_{rs} (B,\Delta ,\alpha,0) \simeq  -2\; \sqrt
\Delta \left( B^2 \; f_{rs} ^{(2)} (\alpha)
+ B^3  \; f_{rs} ^{(3)} \right) \label{qw1}
\end{equation}
where
\begin{eqnarray}\label{f2}
f_{rs} ^{(2)} (\alpha ) &=& \frac 1{\pi} \int
_{0}^{+\infty} \frac{d\nu}{\nu } \; \Phi _{ft} '(\nu ) 
[\Phi _{ft} (\nu) -1 ] - \alpha \; 
\int_{0}^{\infty }dh \left[ \Phi _{cc} (h)
\right] ^2  \\
f_{rs} ^{(3)} &=&  -\frac 1{2\pi} \int
_{0}^{+\infty} \frac{d\nu}{\nu } \; \Phi _{ft} '(\nu ) 
[\Phi _{ft} (\nu) -1 ]^2 \qquad  .
\end{eqnarray}
$f^{(3)}_{rs}$ is clearly positive. Therefore the maximum of $f_{rs}$
is located at $B=0$ if $f^{(2)} _{rs} (\alpha ) \ge 0 $ and at $B>0$
if $f^{(2)} _{rs} (\alpha ) < 0 $. 

It is easy to demonstrate that the threshold $\alpha _c$, determined through 
the condition $f^{(2)} _{rs} (\alpha _c ) =0$ is {\em always} equal to 
unity {\em independently} of the choice of the probability distribution $\Phi$.
To do so, we rewrite the first term on the r.h.s. of (\ref{f2}), that
is $f^{(2)} _{rs} (0 ) $ using the definition (\ref{qw2}) of $\Phi
_{ft}$,
\begin{equation}
f^{(2)} _{rs} (0 ) 
= - \int _{-\infty} ^\infty dx dy \; x\; \Phi (x)  \Phi (y) w(x,y)
\label{qw4}
\end{equation}
with 
\begin{equation}
w(x,y) = \frac 1{2\pi} \int _{-\infty}^ {+\infty} \frac{d\nu}{i \nu }
e^{i \nu x} ( e^{i \nu y} -1 )= \frac 12 \hbox{\rm sign} ( x+y) -
\frac 12 \hbox{\rm sign} (x) \qquad.
\label{qw3}
\end{equation}
Inserting (\ref{qw3}) in (\ref{qw4}), a simple calculation leads to 
\begin{equation}
f^{(2)} _{rs} (0 ) =  \int_{0}^{\infty }dh \left[ \Phi _{cc} (h)
\right] ^2 \qquad , \label{qw7}
\end{equation}
and therefore to the reported result $\alpha _c =1$. Surprisingly,
the variational RS calculation is able to recover the exact threshold
of 2-SAT \cite{alpha2=1} in a very robust manner. Note however, that the
continuous growth of the backbone $B$ above
$\alpha _c$ depends on the choice of $\Phi$.  

\subsubsection{A discontinuous transition: the 3-SAT problem}\label{3satrs}

We now focus on the 3-SAT case ($p=1$). Previous numerical and
analytical studies have shown that spins freeze discontinuously at the
transition \cite{nature,RemiRiccardo1}. 
Thus, we cannot locate the threshold through an expansion
of $f_{rs} (B, \Delta ,\alpha,1)$ as in the 2-SAT case. The full
variational calculation can nevertheless be simplified due to the
following observation. In the SAT phase $(B=0)$, the free-energy is
identically zero. Within a first-order transition scenario, the
threshold will be the value of $\alpha$ at which the free energy of
the UNSAT phase ($B\ne 0$, $\Delta\ne 0$) changes sign to become
thermodynamically stable. The calculation of $\alpha _c$ becomes
simpler once the free-energy (\ref{F2+p1}) is rewritten as
\begin{equation}
f_{rs} (B, \Delta ,\alpha,1) = 2\; \sqrt \Delta
\; B^3 \; \bigg( -s_{rs} (B) + \alpha \; e_{rs} (\Delta) \bigg)
\end{equation}
with
\begin{eqnarray}
s_{rs} (B) &=& \frac 1{\pi B^2} \int
_{0 }^{+\infty} \frac{d\nu}{\nu } \; \Phi _{ft} '(\nu ) \ln
[1-B+B \;\Phi _{ft} (\nu) ] \label{qw6}\\ e_{rs} (\Delta ) 
&=& \int_{0}^{1/{(2 \sqrt \Delta) }}dh \left[ \Phi _{cc} (h) \right] ^3
\qquad . \label{qw5}
\end{eqnarray}
Calling $B_c$ (respectively $\Delta _c$) the argument where $s_{rs} $
(resp. $e_{rs} $) reaches its minimal (resp. maximal) value, we obtain
from (\ref{qw6},\ref{qw5}) the following expression of the threshold
\begin{equation}
\alpha _c = \frac {\min_{B } s_{rs} (B) }{\max_{\Delta } e_{rs} (\Delta) } =
\frac{s_{rs} (B_c)}{e_{rs} (\Delta _c)} \qquad .  
\end{equation}
The maximum of $e_{rs}$ is obviously located at $\Delta _c=0$ whereas the
precise value of $B_c$ depends of the field distribution $\Phi$. 
We list below the results obtained for three different choices.

\begin{itemize}

\item{\em Gaussian distribution:} 
$\qquad \Phi(x) = G_1 (x) \quad , \qquad
B_c \simeq 0.935 \ , \quad \alpha _c \simeq 4.622 \ . $
\item{\em Exponential distribution:}   
$\qquad \Phi(x) = \frac 12 e^{-|x|} \quad , \qquad
B_c \simeq 0.976 \ , \quad \alpha _c \simeq 4.617 \ .$
\item{\em Lorentzian distribution:}
$ \qquad\Phi(x) = \frac 1{\pi \; (1+ x^2 )} \quad , \qquad
B_c \simeq 0.986 \ , \quad \alpha _c \simeq 4.983 \ .$

\end{itemize}
Above the threshold, $B$ and $\Delta$ both increase with $\alpha$ from
their critical values $B_c$ and $\Delta_c (= 0)$. The variational
approach is thus able to reproduce the qualitative picture of the
mixed nature of the phase transition (second order in $\Delta$ and
first order in the backbone size $B$) that emerged from the iterative
RS solution \cite{RemiRiccardo1} and numerics \cite{nature}. This
prediction is quite robust with respect to the choice of $\Phi
(x)$. Even a Lorentzian distribution gives reasonable results for
$\alpha _c$ and $B_c$ though its large-field tail is not physically
sensible, see Section V.B.

  From a quantitative point of view, the above results for the Gaussian
and the exponential case differ from the iterative RS
solution $B_{c}\simeq 0.94$, $\alpha _{c}\simeq 4.60$
\cite{RemiRiccardo1} by a few percent only. However, the latter was
derived through a much less convenient iterative scheme
\cite{RemiRiccardo1}.

\subsubsection{The tricritical point $p_0$}\label{pcrs} 

As we have seen above, the main difference between 2-SAT and 3-SAT
lies in the behaviour of the fraction of frozen spins at the
transition. In other words, the backbone size at threshold vanishes
in the former case ($B_c (p=0)=0$), while it exhibits a discontinuous
jump in the latter case ($B_c (p=1) > 0$).  It is natural to expect
the existence of a tricritical point $p_{0}$ separating continuous
SAT-UNSAT transitions ($p<p_0$) from discontinuous ones ($p>p_0$)
\cite{nature,RemiRiccardo3}.

When $p<p_0$, the transition can be studied through an 
expansion of the free-energy (\ref{F2+p1}) in powers of the backbone
size, see (\ref{qw1}),
\begin{equation}
f_{rs} (B,\Delta ,\alpha,p) \simeq  -2\;\sqrt 
\Delta \left(  B^2 \; f_{rs} ^{(2)}
(\alpha , p) + B^3  \; f_{rs} ^{(3)} (\alpha , p) \right) \label{qw10}
\end{equation}
where, using (\ref{qw4},\ref{qw3},\ref{qw7}),
\begin{eqnarray}
\label{F21}
f_{rs} ^{(2)} (\alpha , p ) &=& \big( 1- (1-p) \; \alpha  \big) \; 
\int_{0}^{\infty }dh \left[ \Phi _{cc} (h)
\right] ^2  \\ \label{F22}
f_{rs} ^{(3)} (\alpha , p )&=&  -\frac {1}{2\pi} \int
_{0}^{+\infty} \frac{d\nu}{\nu } \; \Phi _{ft} '(\nu ) 
[\Phi _{ft} (\nu) -1 ]^2 -\alpha \; p 
\int_{0}^{\infty}dh \left[ \Phi _{cc} (h) \right] ^3\qquad  .
\end{eqnarray}
As long as $f_{rs} ^{(3)} $ remains positive, the threshold is situated
at $\alpha _{c} (2+p) =1/(1-p)$ (\ref{F21}). As in the 2-SAT case, this
result does not depend on the distribution $\Phi$ in (\ref{prsvar}) and
coincides with the rigorous result found in \cite{alpha2+p} for $p
<\frac{2}{5}$.

At a given $p$ and slightly above the threshold, the backbone size 
scales as
\begin{equation} 
B \sim \frac{ \alpha - \alpha _c(2+p)} {f_{rs} ^{(3)} (\alpha _c(2+p) ,
 p )} \qquad ,
\end{equation}
up to a constant multiplicative factor.  The tricritical point $p_0$ can
thus be found through the condition $f_{rs} ^{(3)} (\alpha
_c (2+p) , p)=0$. This statement remains unaffected by the
inclusion of higher order terms in $\Delta$ in the expansion (\ref{qw10}).
The corresponding values of $p_0$ for the three choices of $\Phi$ of
the previous
paragraph are: $p_0 \simeq 0.437$ for the Gaussian distribution,
$p_0 = 3/7 \simeq 0.429$ for the exponential distribution and $p_0
\simeq 0.418$ for the Lorentzian distribution. As expected, these
values are slightly higher than the prediction of the iterative RS
solution, $0.4 \le p_{0} < 0.416$ \cite{RemiRiccardo3}.

We shall now show under some assumptions exposed in Appendix~C that the 
tricritical point is precisely located at
$p_0 = 2/5$. To do so, we proceed in two steps. Firstly, we recall
that the equality $\alpha _c(p) = 1/(1-p)$ for $p \le 0.4$ has been
rigorously demonstrated in \cite{alpha2+p}.  Secondly, using the RS
variational approach, we have seen above that $\alpha _{c} (2+p)
=1/(1-p)$ up to a tricritical $p_0$ which depends of $\Phi$ through
the condition $f_{rs} ^{(3)} (1/(1-p_0 ) , p_0 )=0$.  Consider now two
different Ans{\"a}tze $\Phi ^{(1)}$ and $ \Phi ^{(2)}$ such that the
corresponding tricritical points satisfy $p_0^{(1)} < p_0 ^{(2)}$.
Then, for $p$ in the range $p_0^{(1)} < p < p_0 ^{(2)}$, we have
$\alpha ^{(2)} _c(2+p) = 1/(1-p)$ by definition of $p_0^{(2)}$ and
$\alpha ^{(1)} _c(2+p) < 1/(1-p)$ (the 2-clauses part of the formula
is almost surely satisfiable if and only if $\alpha \cdot (1-p) \le
\alpha _c (2) =1$ giving thus this upper bound to the threshold, see
\cite{nature}). Let us choose $\alpha$ with $\alpha ^{(1)} _c(p) <
\alpha < \alpha ^{(2)} _c(p)$. For Ansatz~2, the free energy $f_{rs}
^{(2)} (\alpha )$ vanishes while within Ansatz~1, $f_{rs}^{(1)}
(\alpha )> 0$. Since the free-energy has to be maximized (see Section
II.A), the first Ansatz has to be preferred to the second
one. Consequently, $p_0$ has to be minimized over the choice of
possible distributions $\Phi$ and an upper bound to the true value of
$p_0$ is provided by the minimal value of $p_0$ within the RS
variational calculation. We show in Appendix \ref{appendix2} that the
latter already equals 2/5.

\subsubsection{Comments on the variational RS calculation}\label{crit}

The main ingredient into our trial variational function is the
separation between the effective fields of order one seen by frozen
spins and the fields of order $T=1/\beta \to 0$ acting on
unfrozen spins. The crucial importance of this separation can be
easily seen {\em a posteriori} with a simple Gaussian Ansatz for $P(h)$,
which amounts to set $B$ to unity in (\ref{F2+p1}). In this case, one finds
$\alpha _{c} \simeq 4.76$ for 3-SAT, a rather high  value. For
2-SAT, the situation is even worse: the predicted value for the
threshold, $\alpha _{c}\simeq 1.7$ is totally wrong while the correct
value $\alpha _c =1$ was successfully obtained by optimizing over $B$.
This result is not surprising: slightly above $\alpha _{c}$, there
are only few constrained spins whereas the Gaussian Ansatz
with $B=1$ and $\Delta>0$ amounts to consider that all spins are frozen. 

Besides its technical simplicity, the variational calculation provides
a better understanding of the transition. While the iterative RS
scheme used in \cite{RemiRiccardo1} was rather involved and the
resulting shape of the field distribution remained unclear, the
two-parameter variational theory presented here stresses unambiguously
the mixed nature of the SAT-UNSAT transition \cite{nature}: of first
order with respect to the backbone size $B$ and continuous with
respect to the intensity $\Delta$ of the effective fields related to
excited configurations. However, from a quantitative point of view,
the predicted value of $B_c$ is much larger than the numerical result
\cite{nature}. This discrepancy stems from an intrinsic weakness of
replica symmetry, which is unable to distinguish between different
kinds of frozen spins (belonging or not to the backbone). As a result,
the parameter $B$ obtained from the variational RS calculation takes
into account all frozen variables and thus overestimates the backbone
size. We shall see in next Section how replica symmetry breaking has
to be introduced to solve this problem.

\subsection{Replica symmetry breaking calculation}\label{RSB}

The variational calculation exposed in Section IV.A as well as the 
iterative replica symmetric solution of \cite{RemiRiccardo1,RemiRiccardo3}
provide qualitative insights into the physical features of the SAT-UNSAT 
transition. From a quantitative point of view, the RS Ansatz however fails 
to predict accurately the threshold of 3-SAT: the estimate $\alpha_ c 
\simeq 4.6$ lies above numerical findings $\alpha _c \simeq 4.25 - 4.30$
\cite{Mitchell,kirk1,nature}. This by itself indicates that a 
replica symmetry broken (RSB) theory of $K$-SAT has to be sought for
\cite{Remi1,RemiRiccardo1,RemiRiccardo3}. Another strong hint is of
course the appearance of RSB in the ground state structure already in
the SAT phase, as discussed in sections III.B and III.C.

\subsubsection{Structure of the RSB field distributions}\label{rsbvar}

A major qualitative weakness of the RS Ansatz underlined in paragraph
IV.A.5 lies in its inability to distinguish spins frozen always in
the same direction (backbone) from spins frozen up and down depending
on the particular ground state cluster. For this reason, the RS
analysis can only predict that the fraction of frozen spins is $\sim
0.93$ but does not tell us the size of the subset that truly belongs to
the backbone of solutions.

Results on the backbone can be obtained within the replica symmetry
breaking analysis. In the latter, the backbone may be defined as the
fraction of frozen spins $S_i$ which do not change direction from
cluster to cluster. The distribution of effective fields $\rho_{i}(h)$
of such a spin has its whole support on the positive (or the
negative) semi-axis only, see Section II.C. Conversely, frozen spins
that do not belong to the backbone can fluctuate from state to state:
their corresponding probability distribution $\rho _{i}(h)$ may extend
over the entire real axis.

On the basis of the previous considerations, we propose the following 
(one step) RSB variational Ansatz,
\begin{eqnarray}\label{p[rho]}
{\cal P}[\rho (h) ]&=&(1-B_{1}-B_{0})\; \delta \bigg[ \rho (h)-\delta 
(h) \bigg] +B_{0}\int_{-\infty }^{+\infty }d\tilde h \; \phi_{0} (\tilde h)
\; \delta \bigg[ \rho (h)-\psi_{0} (h,\tilde h) \bigg]  
 \nonumber \\
&& +B_{1}\int_{-\infty }^{+\infty }d\tilde h\; \phi _{1} (\tilde h )
\; \delta \bigg[ \rho (h)-\psi_{1} (h,\tilde h) \bigg] 
\quad .
\end{eqnarray}
In the above expression, $\delta [.]$ denotes a functional Dirac
distribution \cite{Remi1}.  The first term on the r.h.s. of
(\ref{p[rho]}) is the contribution due to unfrozen spins, whereas the
two other terms include two kinds of frozen spins. $B_0$ gives the
fraction of variables in the backbone. $\psi_{0} (h, \tilde h )$
denotes the distribution of the effective fields $h$ at one site while
fluctuations from site to site are taken into account through $\tilde
h$ and the distribution $\phi _{0} (\tilde h)$
\cite{Beyond,Remi1,Sherr}. Thus, at fixed $\tilde h$, the distribution
$\psi_0(h,\tilde h)$ of $h$ has a support on the semi axis having the
same sign as $\tilde h$.  The last term of (\ref{p[rho]}) is
associated to frozen spins not belonging to the backbone. The
effective field distribution $\psi_{1} (h, \tilde h )$ has therefore
no {\em a priori} restriction on the sign of $h$.

To obtain mathematically tractable expressions, we have made the
following choices for the above field distributions:
\begin{eqnarray}
\label{psi2w}
\phi _{0}(\tilde h) &=&G_{\Delta _0} ( h)\nonumber \\
\psi _{0} (h,\tilde h)&=&\delta (h-\tilde h)\nonumber \\
\phi _{1}(\tilde h) &=&\delta (\tilde h) \nonumber \\
\psi_{1} (h,\tilde h) &=&G_{\Delta _1} ( h-\tilde h) 
\qquad .
\end{eqnarray}
The arbitrary choice for $\phi _{1}$ simply means that
close to the transition the typical value of $\tilde h$ is much
smaller than $h$'s one. Indeed, as in the RS case, effective fields
acting on frozen spins are expected to vanish at the transition
(coming from the high $\alpha$ phase). Ansatz (\ref{psi2w}) is the
simplest one compatible with the sign restriction on the support of
$\psi _0$ and the unbiased distribution of literals in clauses imposing
$P[\rho (h)] = P[\rho (-h)]$.

\subsubsection{Analysis of the transition for 3-SAT}

The trial variational function (\ref{p[rho]}) with (\ref{psi2w}) can 
be plugged into (\ref{crsb})~and~(\ref{freeenergy}). As the temperature 
$T$ is sent to zero, the resulting
variational problem involves four parameters: $B_{1}$, $B_0$,
$r =\sqrt{\Delta _{0}/\Delta _{1}}$ and $\mu  =\beta m
\sqrt{\Delta _{1}}$.  The variances $\Delta _{0}$ and $\Delta _{1}$ of
the fields vanish at the transition (see Appendix C) and enter the
free-energy through the finite ratio $r = \sqrt{\Delta
_{0}/\Delta _{1}}$. Moreover, at very low temperatures, the breakpoint
parameter $m$ naturally behaves as $O(T)$. Since 
the number of states having an excess free energy
$F$ with respect to the lowest lying state scales as
$e^{\beta m F}$ \cite{Beyond}, $\beta m$ keeps finite when $T=0$ not
to spoil RSB effects.
Furthermore, to match the SAT phase ($m=O(1)$, {\em i.e.} $\beta m =
\infty $ ), $\beta m $ has to diverge at the threshold $\alpha _c$ when 
coming from the RSB-UNSAT phase. This divergence
makes $\mu  = \beta m \sqrt{\Delta _{1}}$ finite at the transition.

The variational RSB free-energy is computed in Appendix~D and written
below,  
\begin{eqnarray}\label{rsbfreeenergy}
f _{rsb} (B_{1},B_0,r , \mu  )&=&\frac{1}\mu 
\int_{0}^{+\infty} dx dy \; L(x,y)\left[ 1- B_{1} - B_0 + B_1 e^{-x}
+B_0 e^{-y} \right] \ln \left[ 1- B_{1} - B_0 + B_1 e^{-x} +
B_0 e^{-y} \right] \nonumber \\
&-&\frac{ B_1}\mu   \ln \left( \int _{-\infty}^{+\infty } Dx \;
e^{\;\mu  |x| } \right) - \frac{2 r }{\sqrt{2 \pi}} B_0
- \frac{ \alpha}\mu   B_{1}^{K}  \ln \left( 1-
\frac{2\mu  }{2^{K}H^{K}(0)} \int_{0}^{+\infty } dy\; e^{-2\mu  y}
\; H^{K}(y) \right)\nonumber \\
&+& 2 \alpha \sum_{q=0}^{K-1} {q \choose K} B_{1}^{q}
B_0^{K-q}\int_{0}^{+\infty} dx \left( 
\int_{x / r }^{+\infty} Dy \right) ^{K-q}
\frac{ e^{-2\mu  x} H^{q}(x )}
{2^{q} H(0 )-2\mu  \int_{0}^{x} dy\; e^{-2\mu  y} H^{q}(y) } \quad ,
\end{eqnarray}
where $L(x,y)$, that depends on $r$ and $\mu  $ is the double
inverse Laplace transform of
\begin{equation}\label{G}
{\cal K} (\mu  \sqrt a , \mu  r \sqrt{b} ) =
\int_{0}^{+\infty}dx dy\; e^{-x b  -y a} \; L(x,y) \qquad ,
\end{equation}
with ${\cal K} (a,b)$ defined as 
\begin{equation}
\label{K}
{\cal K} (a,b)=\int_{-\infty}^{+\infty} Dy\ln 
\left[\int_{-\infty}^{+\infty} Dx e^{|ax+by|}
\right]\quad .
\end{equation}
The function $H(x)$ equals
\begin{equation}\label{H}
H(x)=\int_{x}^{+\infty} Dy \; e^{ \; \mu  y }\qquad .
\end{equation}
To compute $f_{rsb}$, we have expanded the logarithm in the first term
of the r.h.s. of (\ref{rsbfreeenergy}) in powers of $B_1$ and $B_0$, using
then the definition (\ref{G}) of $L$ to perform the integrations over
$x$ and $y$. The main difficulty with this procedure is that results
obviously depend on the number $j$ of terms considered in the series
expansion, see Appendix D.  In the simple case $\mu  =0$, we have
checked that the optimal (and $j$-dependent) free-energy $f_{rsb} (j) $
reaches its exact value $f_{rsb}$ with $1/j^{2}$ corrections as $j$
grows. In the general case $\mu  \ne 0$, numerical results support this
scaling: $f_{rsb} (j) = f_{rsb}  + O(1/j^2)$.  For instance
we show in the inset of fig.\ref{fig_transition} the threshold 
$\alpha _{c}(j)$ versus $1/j^{2}$. 
Using this procedure, we have found that the SAT-UNSAT transition
takes place at $\alpha _{c} \simeq 4.480 \pm  0.003$.
This value is still higher than the numerical one, $\alpha _{c}
\simeq 4.25-4.30$ \cite{kirk1,nature}, but definitely improves the RS
result $\alpha _c \simeq 4.60$ and lies below the best known rigorous
upper bounds \cite{bound}. Moreover, at the transition we find
$B_0\simeq 0.13 \pm  0.01$, $B_{1}\simeq 0.79\pm   0.01$, $\mu  \simeq 
0.88\pm  0.02$, and $r \simeq 1.4 \pm  0.1$.

What is the meaning of $\mu$? Consider at a given $\alpha > \alpha
_c$, the clusters corresponding to the ground state energy $E_{GS}$,
{\em i.e.} with the minimal number $=O(N)$ of violated
clauses. Higher-lying clusters $\Gamma$ exist which make slightly more
mistakes: $E_\Gamma = E_{GS} + e _\Gamma$ with $e _\Gamma =
O(1)$. Imagine now that we add $c=O(1)$ new clauses to the instance we
have considered so far. Clearly, the previous ground state assignments
are not necessarily optimal any longer and can be supplanted by
configurations belonging to some clusters $\Gamma$ such that $e_\Gamma
< c$. Therefore, these clusters and the distribution of the related
$e_\Gamma$ are of interest to understand how an instance of $K$-SAT can
adapt in response to some change in the constraints. From a more
quantitative point of view, let us define the linear susceptibility
$\chi _\Gamma$ as the (free-)energy change of the cluster $\Gamma$
when the typical magnitude of the effective fields grows from zero up
to $\sqrt \Delta _1$, divided by $\sqrt \Delta _1$.  Within the RSB
variational approach, the number of quasi-optimal clusters having a
susceptibility equal to $\chi$ scales as
\begin{equation}
{\cal N} (\chi ) \sim e^{\; \mu  \chi } \qquad .
\end{equation} 
The above equation unveils the meaning of the parameter $\mu  $
associated with the breaking of replica symmetry.

\subsubsection{Comments on the replica symmetry breaking solutions}

As stressed in Section II.D, it is crucial to distinguish fields
of the order of one from vanishing fields when $T\to 0$. The
importance of this separation for the RSB solution can be checked
within the variational subspace $B_1 + B_0 =1$ in
(\ref{rsbfreeenergy}), that is discarding unfrozen spins. In this
case, we find $\alpha _{c}\simeq 4.66$ and $B_{1}=1, B_0=0$. This
result is quantitatively and qualitatively erroneous, because the
value of $\alpha _{c}$ is even higher than the value predicted within
the replica symmetric analysis and the fraction of spins belonging to
the backbone is zero. The value of $\alpha _{c}$ can be improved
fixing $B_0=0$ and optimizing over $B_{1}$. In this case we find
$\alpha _{c}\simeq 4.51$, $B_{1}\simeq 0.925$,$\mu  \simeq 0.8$
and there is no backbone.

The backbone is taken into account when relaxing the constraint
$B_0=0$.  The corresponding variational calculation has been exposed
in the previous paragraph.  Let us briefly comment on the results.
First of all, we note that the fraction of frozen spins $B_1 +B_0 \sim
0.92$ changes by a few percents with respect to the RS case. This
value is quite robust and should be quantitatively correct
\cite{remark:frozen}. Conversely, the fraction of spins belonging to
the backbone $B_0 \simeq 0.13$ is underestimated with respect to
numerical findings \cite{nature}, which predict a value of $0.4$ for
small instances.  This probably stems from the choices of the field
distributions (\ref{psi2w}) which break replica symmetry for spins not
belonging to the backbone only. Therefore in the variational treatment
the latter are thermodynamically favoured and the computed fraction of
spins belonging to the backbone is smaller than the true one. Breaking
replica symmetry also for these spins would presumably permit to
obtain better values for $\alpha _{c}$ and $B_0$. This would however
be a hard task due to the technical difficulties arising in the
numerical computation.

To strengthen this intuition, we consider the $\mu  \to 0$ limit of the
RSB free-energy (\ref{rsbfreeenergy}) which amounts to treating the two
kinds of frozen spins on the same footing. The latter becomes
simplified and corresponds to the RS free-energy obtained from the 
following RS field distribution,
\begin{equation}\label{mu=0}
P(h)=(1-B_{1}-B_0) \;\delta (h)+B_{1}\frac{ e^{-h^2 /2 \Delta _{1}}}
{\sqrt{2 \pi \Delta _{1}}} +B_0\frac{ e^{-h^2 /2\Delta _{0}}}{\sqrt{2
\pi \Delta _{0}}}\quad .
\end{equation}
Optimizing $f_{rsb} (B_1 , B_0 , r , \mu  =0)$, we found a
transition at $\alpha _{c}\simeq 4.60$, which quantitatively coincides
with the best know RS solution. The corresponding values of the
variational parameters are $B_{1}\simeq 0.65$, $B_0\simeq 0.29$ and
$r \simeq 0.49$. Once more, the total fraction of frozen spins is
close to $0.94$. However, as conjectured, the relative fraction of
backbone spins increases drastically by a factor of two with respect
to the result of Section IV.B.

\begin{figure}[bt]
\centerline{    \epsfysize=7cm           \epsffile{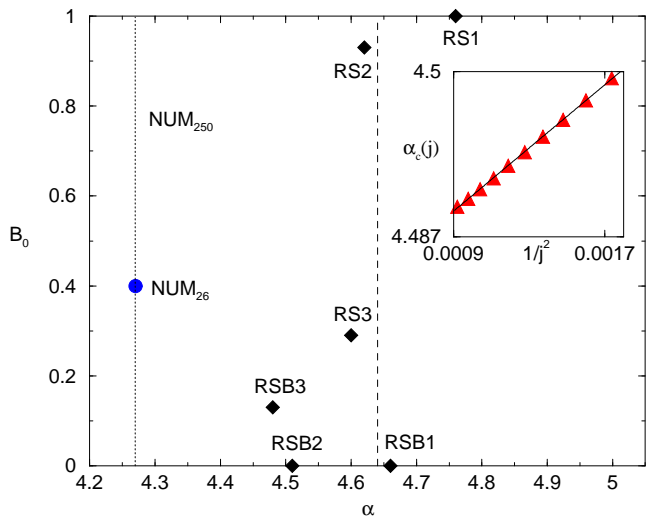}}
\caption{Values of $\alpha _c$ and $B_0$ at the
SAT-UNSAT transition for the different Ans{\"a}tze presented in
Section~IV.  RS1, RS2 and RS3 correspond respectively to the replica
symmetric Ans{\"a}tze with one Gaussian, with one Gaussian and a Dirac
peak, and with two Gaussians and a Dirac peak. RSB1, RSB2 and RSB3 are
their generalizations to the replica symmetry broken case. The dashed
line gives the best known rigorous upper bound on the value of
$\alpha _c$ \protect\cite{bound}. The dotted line and the circle 
respectively show the values of $\alpha _c$ and $B_{0}$ found in 
numerical simulations \protect\cite{nature}.  
Note that the value of $\alpha _{c}$ is more reliable
than the estimate of $B_{0}$ due to the sample sizes used to determine the
former ($N=250$) and the latter ($N=26$). Inset: scaling of $\alpha
_{c} (j)$ as a function of $1/j^{2}$ (where $j$ is the number of terms
considered in the series expansion of the effective entropy
contribution). }
\label{fig_transition}
\end{figure}

The values of $\alpha _c$ and $B_0$ predicted using the above
mentioned Ans{\"a}tze are shown fig.\ref{fig_transition} and
compared to the results of numerical simulations \cite{kirk1,nature}
and the rigorous bounds found in \cite{bound}.

\section{Discussion and conclusion} 

\subsection{$K$-SAT picture arising from the variational calculation}

The variational calculations of the last two sections lead us to
propose the following picture of the $3$-SAT problem.
At very low $\alpha$ each variable $x_i$ is under-constrained, {\it
i.e.}  both SAT instances which result from fixing $x_i$ either to
true or to false are satisfiable with probability one. By adding new
clauses, the number $\alpha$ of constraints per variable is increased
and the solution space shrinks. The latter is made of a single 
cluster without any particular
internal structure. Its diameter $d$ decreases monotonously with the
number of clauses, thus signalling a concentration of the satisfying
assignments  in configuration space, see fig.2. 

\begin{figure}[bt]
\centerline{    \epsfysize=7cm
        \epsffile{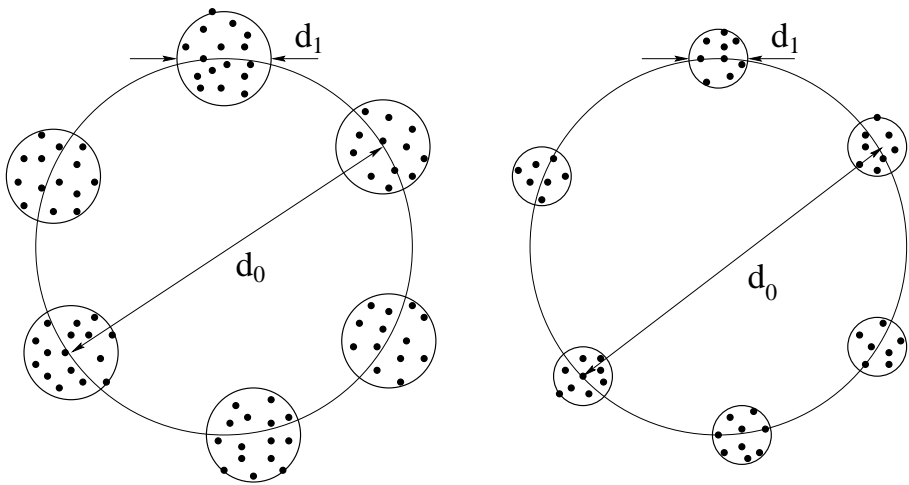}}
\caption{Schematic representation of the solution space structure 
of $3$-SAT for
two values of $\alpha$ with $\alpha_s<\alpha<\alpha_c$ resulting from
the one-step replica symmetry broken variational Ansatz. The solutions
(represented by dots) are organized in clusters. Whereas the 
 distance $d_0$ between the clusters 
remain almost unchanged as $\alpha$ is increased by adding new
clauses, the cluster size $d_1$ decreases quickly.}
\label{fig:structure}
\end{figure}

When $\alpha $ reaches $\alpha_s \simeq 3.96$, the set of all solutions 
continuously breaks up into an exponential number (in $N$) of geometrically 
separated clusters, see fig.\ref{fig:structure} for a schematic 
representation. The instance remains nevertheless satisfiable, and the
variables are still under-constrained. If we further increase the number 
$\alpha N$ of clauses, the typical distance $d_0$ between clusters
remains nearly unchanged. The decrease of the entropy of solutions 
is thus essentially due to the decrease of the average diameter
$d_1$ of the clusters (fig.2).

Increasing $\alpha$ the system becomes unsatisfiable with probability
one at a certain value $\alpha_{c}$, {\it i.e.} it undergoes a
SAT-UNSAT transition.  In the optimal assignments (which minimize the
number of violated clauses), a large fraction of variables
(approximatively $90\%$) becomes over-constrained. The mixed nature of
the SAT-UNSAT transition can be seen explicitely: whereas the fraction
of frozen spins jumps up discontinuously, the effective fields
measuring the strength of the constraints on each variable grow
continuously. Moreover the existence of different clusters of optimal
configurations allows the distinction between two groups of
over-constrained variables. The first group (backbone) contains
variables keeping the same truth value in all optimal
configurations. In the second group, the variables have a
cluster-dependent value. In other words, optimal configurations
corresponding to different thruth values of the second group
variables necessarily belong to distinct clusters and lie at $O(N)$
distances from each other. 

It is important to note that even in the UNSAT regime these frozen
spins coexist with under-constrained variables. These unfrozen
variables lead to a positive entropy at the transition
\cite{remark:frozen}, a behavior which is intrinsically different from
the case of infinite connectivity models \cite{zero.entropy}.

Besides, we should mention
that the actual cluster distribution could be even more complicated,
{\it e.g.} through the existence of clusters of clusters {\it etc.}
The existence of only two typical distances, and thus the
distinction between two kinds of frozen spins is intrinsic to our
one-step replica symmetry broken Ansatz. We nevertheless expect
that the main qualitative features of $3$-SAT are already
captured in our one-step broken variational description. 

Finally, we note that in the $2+p$-SAT case for
$p<p_0=2/5$ the picture arising from the variational calculation is
much simpler.  In fact, $\alpha _s$ and $\alpha_c$ coincide and the
transition from the  under-constrained SAT regime to the
over-constrained UNSAT phase is smooth. The geometrically 
non-trivial intermediate phase does not exist at all.

\subsection{Critical behavior and exponents}

The exact RS saddle-point equation  \cite{RemiRiccardo1}
shows that the probability $P(h)$ that the effective field equals $h
\gg  1$ on a given site is bounded from above by the probability that this
site is connected to $h$ neighbours and then decreases at least  
exponentially with $h$. Combining this observation with the
variational calculation presented in this paper, an investigation of
the optimization equations over $B$ and $H$ 
reveals that slightly above the threshold, the free-energy
exhibits a singularity of the type
\begin{equation}
f_{rs} ( \delta \alpha ) \sim \delta \alpha  \; \cdot 
\left( - \log \delta \alpha \right) ^{-  \eta} \qquad \qquad (
\delta \alpha = \alpha - \alpha _c \ge 0 )\quad .
\label{singu}
\end{equation}
The actual scaling of $\Phi (h)$ at large fields $h$ gives only rise to
logarithmic singularities, e.g. $ \eta =\frac 12$ for a Gaussian
distribution, $ \eta = 1 $ for an exponential one. Note that
equation (\ref{singu}) also holds at the RSB level within the Ansatz
(\ref{p[rho]}).

These predictions can be related to the recent finite-size scaling
(FSS) numerical studies of the $K$-SAT model \cite{kirk1,nature}. Let us
call $E_{GS}(\alpha , N)$ the average ground state energy for a finite
number $N$ of Boolean variables. Close to the threshold, we expect 
the curves of $E_{GS}$ as functions of $\alpha$ obtained for different
sizes $N$ to collapse onto each other when properly rescaled. In other
word, FSS should hold and there should exist some exponents $\nu$ and 
$\gamma$ such that 
\begin{equation}
E_{GS}(\alpha , N) \simeq N^{\gamma} \; {\cal E} \big( N^{1/\nu} \; \delta
\alpha \big) \qquad,
\label{scal45}
\end{equation}
when $\delta\alpha = \alpha - \alpha _c \ll  1$ and $N \gg  1$. $\nu$
characterizes the smaller and smaller width of the transition region 
from the SAT to the UNSAT phase as $N$ grows. It has been numerically
calculated using a Davis-Putnam procedure \cite{kirk1}: $\nu \simeq 3$
for 2+p-SAT as long as $p<p_0$ and $\nu$ decreases for larger $p$ down
to $\simeq 1.5$ for 3-SAT. $\gamma$ can be simply interpreted:
$N^{\gamma}$ is the minimal number of violated clauses at threshold
and thus $\gamma < 1$. The rescaled ground state energy ${\cal E} (y)$
is a monotonously increasing function of its argument:
${\cal E} (y) \to 0$ when $y\to -\infty$ (right boundary of the SAT
phase); ${\cal E}( 0)$ is finite; 
${\cal E} (y) \sim y $ when $y\to \infty$ 
(left boundary of the UNSAT phase). The latter scaling ensures that
$E_{GS}$ grows above the threshold  (that is at fixed
$\delta \alpha$ while $N$ becomes larger and larger) as
\begin{equation}
E_{GS}(\alpha , N) \simeq N^{\gamma} \cdot \; N^{1/\nu}\cdot \; \delta
\alpha  \qquad ,
\label{scal46}
\end{equation}
to coincide with (\ref{singu}) up to logarithmic singularities. 
Imposing that $E_{GS}(\alpha , N)=O(N)$ in the UNSAT regime, identity
(\ref{scal46}) gives the hyperscaling relation
\begin{equation}
\gamma = 1-\frac 1\nu \qquad .
\label{scal47}
\end{equation}
Whereas $\nu$ may be computed for large formulae, involving
thousands of variables, no such powerful method exists so far to
estimate $\gamma$. Therefore,  identity (\ref{scal47}) may be 
precious to derive indirectly $\gamma$ from
the knowledge of $\nu$.

\subsection{Perspectives}

As discussed in Section II.E, the threshold $\alpha _c$ separates $O(T)$
fields (SAT regime) from $O(1)$ ones (UNSAT phase).  This change of
scaling of effective fields is nicely apparent within the variational
calculation presented in Sections~III.A-B and IV.A-B.  Looking at
compatible Ans{\"a}tze for the SAT and the UNSAT phase, the same values
for $\alpha_c$ can be obtained either starting from the SAT phase from
a diverging renormalized variance of the effective fields (which were
assumed to vanish linearly with $T$), or coming from the UNSAT phase
from a vanishing variance of the effective fields (which remained finite
in the limit $T\to 0$). A deeper understanding of the scaling of the
fields in the vicinity of the critical point $\alpha =\alpha _c$,
$T=0$ would be of interest for at least two reasons. 
First, it would allow the calculation of the entropy in the UNSAT phase,
which has been out of reach yet. Secondly, the structure of the
invariant measure $P(h)$ could be studied carefully to gain some 
information  on its singularities, its fractal 
structure,~etc. \cite{zagrebnov}.
At finite but low temperatures, one might expect that the
support of $P(h)$ includes different regions corresponding to
different scalings with $T$, {\it i.e.} to distinct physical phenomena
coexisting in the model. This potential richness of the order
parameter cannot be present in infinite-connectivity spin
glasses and could give rise to new properties at the mean-field level.

Further work is clearly required to confirm the geometrical picture of
the space of solutions sketched in Section V.A. From a numerical point
of view, an analysis of the distances between solutions would be of
interest to check the existence of a non trivial (non necessarily
bimodal) distribution for $d$. It would be also worth trying to
improve our analytical approach by using richer trial field
distributions in the RSB calculations. However, the most promising
route is probably to attempt to use the information presently
available on the optimal (and quasi-optimal, see Section IV.B)
assignments of $K$-SAT to understand the drastic change of behaviour
of algorithms close to the threshold.

\vskip .5cm
{\bf Acknowledgements:} We are grateful to O.~Dubois, S.~Kirkpatrick
and R.~Zecchina for useful discussions, and J. Berg for carefully
reading the manuscript. M.W. acknowledges financial support by the 
German Academic Exchange Service (DAAD).

\vskip .5cm
{\bf Note added in proofs:} After submission of this paper we have been
 aware of a new rigorous upper bound for 3-SAT, $\alpha _{c}\le 4.506$
proven by O. Dubois. This upper bound lies slightly above our RSB result.
\appendix

\section{RSB free-energy in the SAT phase}
 
In this appendix we show how the one-step replica symmetry broken
ground state entropy is calculated. We start from (\ref{freeenergy}),
perform the limit $\beta \to\infty$:
\begin{equation}
\label{A:entropy}
s = \lim_{n\to 0} -\frac{1}{n}\sum_{\vec{\sigma}} 
c(\vec{\sigma}) \ln c(\vec{\sigma})  
+\frac{\alpha}{n} \ln\left[ \sum_{\vec{\sigma}_1,...,\vec{\sigma}_K} 
c(\vec{\sigma}_1)\cdots c(\vec{\sigma}_K) \prod_{a=1}^n \left( 1-
\prod_{a=1}^n \delta_{\sigma_l^a,1} \right)\right]
\end{equation} 
and plug in  Ansatz (\ref{AnsatzrsB}), that is
\begin{equation}
\label{crsB}
c(\vec{\sigma}) = \int_{-\infty}^\infty dz\  G_{\Delta_0}(z)
\prod_{a=1}^{n/m} \frac{\int d\tilde{z} G_{\Delta_1}(\tilde{z}-z)
\exp\left\{ \tilde{z} \sum_{b=(a-1)m+1}^{am} \sigma ^{b}\right\}}{
\int d\tilde{z} G_{\Delta_1}(\tilde{z}-z) (2 \cosh \tilde{z})^m}
\quad .
\end{equation}
We will calculate both terms on the r.h.s. of (\ref{A:entropy})
separately. We start with the effective entropic term and follow
closely the analytical continuation scheme proposed in \cite{Remi1},
\begin{eqnarray}
  \label{A:entr}
-\lim_{n\to 0} \frac{1}{n}\sum_{\vec{\sigma}} c(\vec{\sigma}) \ln
c(\vec{\sigma})
&=& -\int {\cal D}\hat{\nu}{\cal D}\nu\ \exp\left\{
-i\int_{-\infty}^\infty
dy\ \hat{\nu}(y)\ \nu(y) \right\} c[i\nu]\ln c[i\nu] \nonumber\\
&& \ \ \ \ \ \ 
   \frac{1}{m} \ln\left\{\int_{-\infty}^\infty \frac{dx\ dy}{2\pi}
e^{-ixy}(2\cosh x)^m \exp \hat{\nu}(iy) \right\}
\end{eqnarray}
with
\begin{equation}
\label{A:op}
c[\nu]  = \int {\cal D}\rho {\cal P}[\rho]
\exp\left\{\int_{-\infty}^\infty
dy\ \nu(y) \ \ln\left[ \int_{-\infty}^\infty dh\ \rho(h) \frac{e^{\beta
hy}}{(2\cosh\beta h)^m} \right]\right\}\ .
\end{equation}
Note that this form does not depend on $K$, {\it i.e.} 
on the length of the 
clauses, and the following calculations are consequently valid for any
$K$. With Ansatz (\ref{AnsatzrsB}), the last expression depends on
$\nu(y)$ only through its first three moments,
\begin{equation}
  \label{A:cnu}
c[\nu] = c(\nu_0,\nu_1,\nu_2)
= \int_{-\infty}^\infty dz\  G_{\Delta_0}(z) \exp\left\{
-\nu_0 \ \ln \int_{-\infty}^\infty 
d\tilde{z}\ G_{\Delta_1}(\tilde{z}-z)(2\cosh\tilde{z})^m
+ z\nu_1 + \Delta_1 \nu_2 \right\}
\end{equation}
with
$\nu_l = \int_{-\infty}^\infty dy\ \nu(y)\ y^l/l!$ $(l=0,1,2)$. 
The effective entropic part can now be calculated according to 
(\ref{A:entr}) if we introduce the series expansion 
$\hat{\nu}(y) = \sum_{l=0}^\infty \nu _l y^l /l!$. Using
$\int dy\ \nu(y)\hat{\nu}(y) = \sum _{l=0}^\infty \nu_l\hat{\nu}_l$
the integrals over the $\nu_l$ and $\hat{\nu}_l$ with $l\geq 3$ can be 
executed trivially; $\nu_l$ does not exist outside the first exponential 
leading to a Dirac-function in  $\hat{\nu}_l$, which vanishes
consequently.
Thus we obtain
\begin{eqnarray}
  \label{A:entrop}
-\lim_{n\to 0} \frac{1}{n}\sum_{\vec{\sigma}} c(\vec{\sigma}) \ln
c(\vec{\sigma})
&=& - \int \prod_{l=0}^2 \left(\frac{d\nu_l\ d\hat{\nu}_l}{2\pi}
e^{-i \nu_l\hat{\nu}_l} \right) c(i\nu_0,i\nu_1,i\nu_2) \ln 
c(i\nu_0,i\nu_1,i\nu_2) \nonumber\\
&&\ \ \ \     \frac{1}{m} \ln \left[ \int \frac{dx\ dy}{2\pi} e^{-ixy} 
(2\cosh x)^m \exp\{ \hat{\nu}_0+i\hat{\nu}_1 y - \frac{1}{2}\hat{\nu}_2
y^2 \} \right]\nonumber\\
&=& S_0 +S_1 \qquad ,
\end{eqnarray}
where
\begin{eqnarray}
  \label{A:S0}
  S_0 &=& -\frac{1}{m} \int \frac{d\nu_0\ d\hat{\nu}_0}{2\pi}
e^{-i \nu_0\hat{\nu}_0}\ \hat{\nu}_0\ c(i\nu_0,0,0) \ln c(i\nu_0,0,0)
\nonumber\\
&=& +\frac{1}{m} \int dz\ G_{\Delta_0}(z) \ \ln \int d\tilde{z}\  
G_{\Delta_1}(\tilde{z}-z) \left( 2\cosh \tilde{z} \right)^m
\end{eqnarray}
and
\begin{eqnarray}
  \label{A:S1}
S_1 &=& -\frac{1}{m} \int \frac{d\nu_1\ d\hat{\nu}_1\ d\nu_2\
d\hat{\nu}_2}{
(2\pi)^2}e^{-i(\nu_1\hat{\nu}_1+\nu_2\hat{\nu}_2)}\ c(0,i\nu_1,i\nu_2) 
\ln c(0,i\nu_1,i\nu_2) \nonumber\\
&&\ \ \ \  \times     \ln \left[ \int \frac{dx\ dy}{2\pi} e^{-ixy} 
(2\cosh x)^m \exp\{ i\hat{\nu}_1 y - \frac{1}{2}\hat{\nu}_2
y^2 \} \right]\nonumber\\
&=& -\frac{1}{m} \left( \Delta_0\frac{\partial}{\partial\Delta_0}
+ \Delta_1\frac{\partial}{\partial\Delta_1} \right)
\int dz\ G_{\Delta_0}(z) \ \ln \int d\tilde{z}\  
G_{\Delta_1}(\tilde{z}-z) \left( 2\cosh \tilde{z} \right)^m
\end{eqnarray}
By calculating the derivatives and rescaling the integration variables
we finally find the corresponding contribution in (\ref{entropyrsB}). 

Let us now calculate the effective energy, {\em i.e.} the
explicitly $\alpha$-dependent contribution in (\ref{entropyrsB}) starting
from the last term in (\ref{A:entropy}) by plugging
in (\ref{crsb}). The sum over the replicated spin
variables can easily be carried out. This directly gives
\begin{equation}
 E = -\lim_{n\to 0} \frac{\alpha}{n} \ln\left[
\int\prod_{l=1}^K ( dz_l\ G_{\Delta_0}(z_l) )
\left(\frac{
\int\prod_{l=1}^K ( d\tilde{z}_l\ G_{\Delta_0}(\tilde{z}_l-z_l) ) 
\left(\prod_{l} 2\cosh\tilde{z}_l -\prod_{l}e^{\tilde{z}_l}
\right)^m}{
\int\prod_{l=1}^K ( d\tilde{z}_l\ G_{\Delta_0}(\tilde{z}_l-z_l) 
( 2\cosh\tilde{z}_l)^m)
}\right)^{\frac{n}{m}}\right]
\end{equation}
In the limit $n\to 0$ and after a rescaling and translation of the
integration variables to normally distributed Gaussian variables, we
find the corresponding expression in (\ref{entropyrsB}).


\section{RS Gaussian Ansatz for the SAT-UNSAT transition}

In this appendix, we compute the replica symmetric variational
free-energy for a Gaussian distribution $\Phi (x) =
G_1(x)$. Using (\ref{F2+p1}), we straightforwardly obtain for the
$K$-SAT problem
\begin{equation}
f_{rs} ^{Gauss} (B,\Delta,\alpha, K) = 2 B \sqrt 
\Delta \left\{  \int _{-\infty} ^\infty
\frac{dx}{2\pi} e^{-x^2/2} \ln \left( 1- B + B\; e^{-x^2/2}
\right) + \alpha\; B^{K -1} \int _{0} ^{1/(2\sqrt \Delta )} dx \left[ 
\int _{x} ^\infty Dy \right]^{K}
\right\}\ .
\label{fgrs}
\end{equation}
The corresponding free-energy for the $2+p$-SAT model can be easily
obtained by a linear combination of expression(\ref{fgrs})
for $K=2$ (with weight $1-p$) and $K=3$ (with weight $p$).

However for completeness we give a derivation of  free-energy 
(\ref{fgrs}) from (\ref{freeenergy}) without any reference to 
\cite{Remi1}. The order parameter $c(\vec{\sigma })$ reduces to
\begin{equation}\label{crsgaus}
c(\vec{\sigma })=\frac{1-B}{2^{n}}+B\int_{-\infty }^{+\infty }
Dx \prod _{a=1}^{n}
\frac{e^{\beta \sqrt \Delta x \sigma ^{a} }} {2\cosh \beta \sqrt \Delta x}\qquad .
\end{equation}
The first term on the r.h.s. of (\ref{freeenergy}) may then be written
in the limit $n\to 0$ and $\beta \to \infty$ as 
\begin{equation}\label{eq1entrors}
\frac{1}{\beta n}\sum_{\vec{\sigma }} c(\vec{\sigma }) \ln c(\vec{\sigma
})= \frac{1}{\beta n}\frac{\partial}{\partial l}\sum_{\vec{\sigma }} 
\left. c(\vec{\sigma })^{l}\right|_{l=1} =-\frac{\sqrt \Delta}{\sqrt \pi } \;
\frac{\partial}{\partial l}\left. \left[
\sum_{j=0}^{l}{ l \choose j}(1-B)^{l-j}B^{j}(j
-\sqrt{j})  \right]\right|_{l=1} \qquad .
\end{equation}
The difficulty in the computation of the entropic contribution is the
analytic continuation in $l$. For the r.h.s. of (\ref{eq1entrors}),
this can be easily achieved through the relation
\begin{equation}\label{exp}
j-\sqrt{j}=j\left(1-\int_{-\infty }^{+\infty } \frac{dx}{\sqrt{2 \pi }}
\; e^{-x^{2}j}\right) \qquad . 
\end{equation}
Using (\ref{exp}), the sum over $l$ in (\ref{eq1entrors}) can be
carried out and expression (\ref{fgrs}) recovered.  The last term of
(\ref{frs}) is found when inserting the RS expression of
$c(\vec{\sigma })$ in (\ref{freeenergy}) and dividing by $n$.  Using
the variational expression (\ref{p(h)}) of $P(h)$, this term reduces
to $\alpha B^{K} \sqrt \Delta \int_{0}^{+\infty}Dx _1 \cdots Dx_{K}
\min \left[1/\sqrt \Delta,2 x_1, \dots , 2 x_{K} \right]$ when $\beta
\to \infty$.  It is easy to check that the above expression coincides
with the last term of (\ref{fgrs}).


\section{Variational upper bound to $p_0$}\label{appendix2}

For a given distribution $\Phi$, the tricritical point $p_0$ is
obtained through the condition $f_{rs} ^{(3)} (1/(1-p_0 )
, p_0 )=0$. Using (\ref{F22}), we obtain
\begin{equation} \label{rt5}
p_0 [\Phi ] = \frac {A[\Phi ]}{A[\Phi ]+B[\Phi ]} \qquad ,
\end{equation}
where 
\begin{eqnarray}
\label{rdephi}
A [\Phi ] &=& - \frac 1{2\pi} \int
_{0}^{+\infty} \frac{d\nu}{\nu } \; \Phi _{ft} '(\nu ) 
[\Phi _{ft} (\nu) -1 ]^2 \\
B[\Phi ] &=& \int_{0}^{\infty}dh \left[ \Phi _{cc} (h) \right]
^3 \qquad.
\end{eqnarray}
The extremization condition of $p_0$ with respect to the even
distribution $\Phi$ may thus be written as 
\begin{equation}
\label{col}
\frac{\delta p_0 }{\delta \Phi (x) } 
[\Phi  ] =\lambda ,\qquad \forall x \ge 0 ,
\end{equation}
where $\lambda$ is a Lagrange multiplier ensuring the normalization of
$\Phi $. Equation (\ref{col}) involves the functional
derivatives of $A$ and $B$,
\begin{eqnarray} \label{ccol1}
\frac{\delta A }{\delta \Phi (x) } [\Phi ] &=& 2\;x \; \int _{-\infty}
^\infty dy\; \Phi(y) \big[ \Phi _{cc} (x+y) - \Phi _{cc} (x) \big] + 2
\; \int _{-\infty} ^\infty dy\; y\; \Phi(y) \big[ \Phi _{cc} (x-y) -
\theta (y-x) \big] \\ 
\label{ccol2}
\frac{\delta B }{\delta \Phi (x) } [\Phi ] &=& 3
\; \int _{0} ^x dy\; \big[ \Phi _{cc} ( y ) \big]^2 \qquad ,
\end{eqnarray}
where $\theta (\cdot )$ denotes the Heaviside function.
By subtracting the values of the functional derivatives of $p_0$
(\ref{col}) in $x=0$ and $x=\infty$, the Lagrange multiplier $\lambda
$ disappears. We obtain $B[\Phi] / A[\Phi] = 3/2$ and therefore from
(\ref{rt5}),
\begin{equation} \label{resuphinal}
\min _{\Phi}\; p_0 [\Phi ] = \frac 25 \qquad .
\end{equation}
The determination of the optimal distribution $\Phi$, although of
interest, see Section V.B would be more difficult. Note that we have
implicitely assumed in the functional differentiations 
(\ref{col},\ref{ccol1},\ref{ccol2})   
that $\Phi$ included no Dirac distributions. The value
of $p_0=2/5$ directly comes from this hypothesis as shown in
\cite{RemiRiccardo3}.


\section{RSB  free-energy for the SAT-UNSAT transition}\label{appendix3}

Within the RSB Ansatz (\ref{p[rho]}), the order parameter
 $c (\vec{\sigma})$  reduces to:
\begin{equation}\label{rsbcsigma}
c(\vec{\sigma})=\frac{1-B_1 -B_0}{2^n}+B_1 \prod_{b=1}^{\frac{n}{m}}
\frac{\int_{-\infty}^{+\infty} Dh \; e^{\beta h \sqrt{\Delta _1} s_b}}{
\int_{-\infty}^{+\infty} Dh (2\cosh \beta h \sqrt{\Delta _1})^m}
+B_0\int_{-\infty }^{+\infty }
Dh\prod _{a=1}^{n}\frac{e^{\beta h \sigma ^{a} \sqrt{\Delta _0}}}
{2\cosh \beta h \sqrt{\Delta _0}} \quad . 
\end{equation}
In the following, we compute the effective entropy contribution
by taking the derivative of $\sum_{\vec{\sigma }} c(\vec{\sigma})^{l}$
with respect to $l$, see (\ref{eq1entrors}). For $\beta \to \infty$, 
we find
\begin{eqnarray}
S&=&\frac{ \sqrt{\Delta _{1}}}\mu  \frac{\partial}{\partial l}\left\{
\sum_{p=0}^l {l \choose p}(1-B_1 -B_0)^{l-p}\sum^{p}_{q=0}
{p \choose q} B_1
^{q} B_0 ^{p-q} \int_{-\infty}^{+\infty}
 Dh_{q+1}\cdots Dh_{p}\right.\nonumber \\
&&\left[\ln\left(\int_{-\infty}^{+\infty} Dh_1\cdots Dh_{q}e^{\; \mu 
|(h_1 +\cdots  h_{q})+(h_{q+1}+\cdots h_p)r|}
\right)\right. \nonumber\\
&&\left.\left. \left.-q\ln \left(\int_{-\infty}^{+\infty} Dh \; e^{2\mu  |h|} 
\right)-\frac\mu  {\sqrt{\Delta_1}}(p-q)\int_{-\infty}^{+\infty}
 Dh|h|\sqrt{\Delta _0}
\right]
\right\}\right| _{l=1}
\quad , \label{hsy}
\end{eqnarray}
where $\mu  =m \beta \sqrt{\Delta _{1}}$ and $r=
\sqrt{\Delta _0/\Delta _1}$. 
As in the replica-symmetric computation of appendix B, the
main difficulty is the analytic continuation in $l$.  For the last two
terms in (\ref{rsbcsigma}) the sum over $l$ can be performed explicitly,
therefore the analytic continuation can be trivially performed. It is
easy to check that the contribution to $S$ of these two terms leads to
the second and the third terms of (\ref{rsbfreeenergy}) multiplied by
$\sqrt{\Delta _{1}}$.  For the first term of (\ref{hsy}) the
analytic continuation in $l$ is more tricky. First of all using the
convolution properties of Gaussian functions this term is reduced to
\begin{equation}\label{series2}
S_{I}=\frac{ \sqrt{\Delta _{1}}}\mu  \left.
\frac{\partial}{\partial l}\left\{
\sum_{p=0}^l {l \choose p}(1-B_1 -B_0)^{l-p}\sum^{p}_{q=0}
{p \choose q} B_1
^{q} B_0 ^{p-q}{\cal K} (\mu  \sqrt{q},\mu  r\sqrt{ p-q})\right\}
\right|_{l=1}\quad , 
\end{equation}
where the function ${\cal K} (a,b)$ has been defined in (\ref{K}).Then
the analytic continuation in $l$ can be achieved using the function $L
(x,y)$ defined in (\ref{G},\ref{K}) and the first term of
(\ref{rsbfreeenergy}) (multiplied by $\sqrt{\Delta _{1}}$) is
recovered. Although written in a compact way, the resulting $S_I$ is
not very useful for numerical purposes. We have rather use the
equivalent expression
\begin{eqnarray}\label{sII2}
S_{I}&=-&\frac{ \sqrt{\Delta _{1}}}
\mu  \sum_{l=1}^{+\infty}\frac{1}{l}
\sum_{p=0}^l (-1)^{p} {l \choose p}(B_1 +B_0)^{l-p}\sum^{p}_{q=0} 
{p\choose q} B_1 ^{q} B_0 ^{p-q}
\left[ B_{1} {\cal K} (\mu  \sqrt{  q+1},\mu  r \sqrt{ p-q})
\right.\\
&& \left. +B_0
{\cal K} (\mu  \sqrt{q},\mu  r\sqrt{ p-q+1})+
(1-B_{1}-B_0) {\cal K} (\mu  \sqrt{q},\mu  r \sqrt{ p-q})
\right]\quad .\nonumber
\end{eqnarray}
 
We now turn to the effective energy contribution which reads \cite{Remi1},
\begin{eqnarray}
E &=& -\frac{\alpha}{ \beta}
\int _V {\cal D} \rho _1 \ldots {\cal D} \rho _K \; {\cal P}[\rho _1 ]
\ldots  {\cal P}[\rho _K ]  \nonumber \\
&\times  & \frac 1m \ln \left[ \int
_{-\infty}^{+\infty}
 dh_1 \ldots dh_K \rho _1(h_1) 
\ldots \rho _K (h_K) \left( 1 + (e^{-\beta} -1) \frac{ e^{\beta \sum
_{j=1}^ K h_j}}{ \prod _{j=1}^K 2 \cosh \beta h_j } \right)
^m \right] \quad . 
\label{enerksat}
\end{eqnarray}
Plugging the trial variational functional (\ref{p[rho]}) 
into (\ref{enerksat}), we find for $\beta \to \infty$,
\begin{eqnarray}\label{energetic2}
E&=& -\frac{\alpha \sqrt{\Delta _{1}}}\mu 
\sum_{q=0}^{K} {K \choose q} B_{1}^{q}B_0^{K-q}
\int_{0}^{+\infty}Dh_{q+1}\cdots Dh_{K}\left\{ -q\ln \int_{-\infty }^{+\infty}
Dh e^{\mu  |h|}\right.\nonumber \\
&+& \left.\ln\left[\int_{-\infty }^{+\infty } Dh_{1}e^{\mu 
|h_{1}|}\cdots Dh_{q}e^{\mu  |h_{q}|}\left( 1+
\prod _{i=1}^{q} \theta (h_{i})[ e^{-\mu  \; \min (1 / \sqrt{\Delta _{1}},
2h_{1},\dots ,2h_{q} ,2r h_{q+1} ,\dots , 2r h_{K} ) }-1 ]\right) 
\right]  \right\} \quad ,
\end{eqnarray}
where $\theta (h)$ is the Heaviside function.
First of all we focus on the $q=K$ term, which is the only one that
 does not vanish for $B_{1}=1$. In this case a simple integration by
 parts leads to
\begin{equation}\label{firstterm}
-\frac{\alpha B_{1}^{K} \sqrt{\Delta_{1}}}\mu  \ln \left(1-
\frac{2\mu  }{2^{K}H^{K}(0)} \int_{0}^{1/(2 \sqrt 
\Delta _{1})} e^{-2\mu  h}H^{K}(h)
\right)\quad,
\end{equation}
where the function $H$ has been defined in (\ref{H}).
The other terms in the sum over $q$ lead to two different contributions.
For $h_{q+1},\dots ,h_{K}>1/2
\sqrt{\Delta _{0}}$ by an integration by parts 
the integrals in the {\em q}th term reduce to:
\begin{equation}\label{energeticdif1}
\left(\int_{\frac{1}{2\sqrt{\Delta _{0}}}}^{+\infty }
Dh\right)^{K-q}\ln \left(1-
\frac{2\mu  }{2^{q}H^{q}(0)} \int_{0}^{\frac{1}{2\sqrt{\Delta
_{1}}}}e^{-2\mu  h}H^{q}(h)
\right) \quad .
\end{equation}
On the other hand if there is at least one field among $h_{q+1},\dots
,h_{K}$ which is smaller than $1/2 \sqrt{\Delta _{0}}$. After a little
algebra we find that the integrals in the {\em q}th term reduce
to
\begin{equation}\label{energeticdif2}
(K-q)\int_{0}^{\frac{1}{2\sqrt{\Delta_{0}}}}Dh 
\left(\int_{h}^{+\infty }
Dh\right)^{K-q-1}
\ln \left[ 1-\frac{2\mu  }{2^{q}H^{q}(0)} \int_{0}^{
r h }dh'e^{-2\mu  h'
}H^{q}(h')\right]
 \quad .
\end{equation}
Collecting (\ref{energeticdif1}) and (\ref{energeticdif2}) 
and integrating by parts, we find that the
 sum over $q$ (for $q\neq K$) reduces to
\begin{equation}\label{energefin}
2 \alpha \sqrt{\Delta _{1}} \sum_{q=0}^{K-1}{K\choose q}B_{1}^{q}
B_0^{K-q}\int_{0}^{\frac{1}{2\sqrt{\Delta _{0}}}}dh 
\left(\int_{h/r }^{+\infty }
Dh' \right)^{K-q}
\frac{e^{-2\mu  h}H^{q}(h)}
{2^{q} H(0)-2\mu  \int_{0}^{h}dh' e^{-2\mu  h'} H^{q}(h')}
\quad .
\end{equation}
Finally gathering (\ref{energefin}) and (\ref{firstterm}) one
obtain the final form of the effective energy part.

It is easy to verify that both $\Delta_{1}$ and $\Delta_{0}$ vanish at
the transition while the ratio $r=\sqrt{\Delta_{0}/\Delta_{1}}$ has a
non trivial value.  Dividing the variational free energy by
$\sqrt{\Delta_{1}}$ the entropic contribution depends on $\Delta_{1}$
and $\Delta_{0}$ through $r$ alone. Therefore as in the replica
symmetric analysis the optimization on $\Delta_{1}$ has to be
performed for the energetic contribution only. This procedure leads to
$\Delta_{1}=\Delta_{0}=0$ at the transition. As a consequence within
the variational approach the analysis of the SAT-UNSAT transition
reduces to the study of the variational free energy
(\ref{rsbfreeenergy}).



\begin{thebibliography}{99}

\bibitem{trans}
P. Cheeseman, B. Kanefsky, and W. M. Taylor, in {\it Proc. 13th
Int. Joint Conf. Artif. Intell. (IJCAI-91)},
(eds. J. Mylopoulos and K. Reiter), 331 (Morgan Kaufmann, San Mateo, 
California, 1991).\\
D. Mitchell, B. Selman, and H. Levesque, in {\it Proc. 10th
Natl. Conf. on Artif. Intell. (AAAI-92)}, 440 (AAAI Press/MIT Press, 
Cambridge, Massachusetts, 1992).\\
T. Hogg, B. A. Huberman, and C. Williams (eds.), Frontiers in problem
solving: phase transitions and complexity.
{\it Artif. Intell.} {\bf 81} (I+II) (1996).

\bibitem{GaJo} M. R. Garey and D. S. Johnson, {\it Computers and
Intractability, A Guide to the Theory of NP-Completeness}, (Freeman, 
San Francisco, 1979).

\bibitem{Papa} C. Papadimitriou, {\it Computational Complexity},
(Addison-Wesley, Readings, 1994)

\bibitem{nature}
R. Monasson, R. Zecchina, S. Kirkpatrick, B. Selman, L. Troyansky,
{\it Nature} {\bf 400}, 133 (1999).
                                
\bibitem{alpha2=1}
A. Goerdt, in {\em Proc. 7th Int. Symp. on Mathematical Foundations
 of Computer Science}, 264 (1992).\\
V. Chantal and B. Reed, in {\em Proc. 33rd IEEE Symp. on Foundations
 of Computer Science}, 620 (IEEE Comp. Soc. Press, New York, 1992).

\bibitem{Mitchell}
J.M. Crawford, L.D. Auton, in {\it Proc. 11th
Natl. Conf. on Artif. Intell. (AAAI-93)}, 21 (AAAI Press, 
Menlo Park, California, 1993).

\bibitem{kirk1}
B. Selman, S. Kirkpatrick, {\it Science} {\bf 264}, 1297 (1994).

\bibitem{dedominicis}
C. De Dominicis and P. Mottishaw, {\em J. Phys.} {\bf A20}, L375 (1987).

\bibitem{RemiRiccardo1}
R. Monasson and R. Zecchina, {\em Phys. Rev. Lett.} {\bf 76}, 3881
(1996), {\em Phys. Rev. E {\bf 56}, 1357 (1997)}.

\bibitem{RemiRiccardo3}
R. Monasson and R. Zecchina, {\em J. Phys.} {\bf A31}, 9209 (1998).

\bibitem{Remi1}
R. Monasson, {\em J. Phys. A {\bf 31}, 513 (1998)}.
 
\bibitem{self}
A.Z, Broder, A.M. Frieze and E. Upfal, 
in {\sl Proc. 4th Annual ACM--SIAM Symp. on Discrete
Algorithms}, 322 (1993).

\bibitem{Beyond}
M. M{\'e}zard, G. Parisi and M. Virasoro, {\em 
Spin Glass Theory and Beyond}, (World Scientific, Singapore, 1987).

\bibitem{pspindedom}
This is a general problem, which arises in the study of disordered
spin systems with finite connectivity. So far, an exact replica symmetry
breaking solution has been found for the diluted p-spin model when
$p\rightarrow +\infty $ only \cite{dedominicis}.

\bibitem{varia9}
E.I. Shaknovich and A.M. Gutin {\em Europhys. Lett. {\bf 8}, 327 (1991)}.\\
 M. M{\'e}zard and G. Parisi {\em 
J. Phys. A {\bf 23}, L-1229 (1990)}; {\em J. Physique {\bf I} 1, 809(1991)}.

\bibitem{Ga} E. Gardner, {\it Nucl. Phys.} {\bf B257}, 747 (1985).\\
A. Crisanti and H.J. Sommers, {\it Z. Phys.} {\bf B87}, 341 (1992).

\bibitem{Sv} P. Svenson and M. G. Nordahl, {\it Phys. Rev.} 
{\bf E59}, 3983 (1999).

\bibitem{cuku} L. F. Cugliandolo and J. Kurchan, {\it Phys. Rev. Lett.}
{\bf 71}, 173 (1993).\\
J. P. Bouchaud, L. F. Cugliandolo, J. Kurchan and M. M{\'e}zard,
in {\it Spin Glasses and Random Fields}, ed. A. P. Young,
(World Scientific, Singapore, 1998).

\bibitem{numerics} The sixfold numerical integration is somewhat
subtle. Even if the integrand shows an asymtotically exponential
behavior we have used the Gauss-Hermite quadrature. This works well
only if $\sqrt{\Delta_1}\tilde{h}$ is small compared to $\tilde{h}^2$
outside the region covered by the abscissas used in the quadrature.
However, this is fulfilled in our case because the interesting
values of $\Delta_1$ are of order one. We have cross-checked our
results by using a Monte-Carlo integration and a series expansion
of the integrand allowing a reduction of the number of integrations to
2 at the cost of a double series. In the case of $\partial_m s_{1RSB}
(m=1,\Delta_0=\Delta_{RS},\Delta_1)$ the results where precise up to
$\Delta_1\approx 20$, then a systematic deviation showed up.

\bibitem{Mo2} R. Monasson, {\it Phys. Rev. Lett.} {\bf 75}, 2847 
(1995).

\bibitem{FrPa} S. Franz and G. Parisi, {\it J. Phys.} {\bf I5}, 1401
(1995). 

\bibitem{MoKa} R. Monasson and D. O'Kane, {\it Europhys. Lett.}
{\bf 27}, 85 (1994).\\
S. Cocco, {\it Tesi di Laurea}, Roma (1995).\\
S. Cocco, R. Monasson and R. Zecchina, {\it Phys. Rev.}
{\bf E54}, 717 (1996).\\
M. Weigt and A. Engel, {\it Phys. Rev.} {\bf E55}, 4552 (1997).

\bibitem{Ni} Th. M. Nieuwenhuizen, {\it J. Phys. I (France)} {\bf 4},
1819 (1996).\\
M. E. Ferrero, G. Parisi, and P. Ranieri, {\it J. Phys.} {\bf A29},
L569 (1996).

\bibitem{DoKoTe} T. Temesvari, C. de Dominicis, and I. Kondor, 
{\it J. Phys.} {\bf A27}, 7569 (1994).

\bibitem{alpha2+p}
D. Achilioptas, L.M. Kirousis, E. Kranakis and D. Krinzac, in 
{\em Proc. of RALCOM 97}, 1 (1997).

\bibitem{Sherr}
K.Y.M. Wong, D. Sherrington, {\em J. Phys.} {\bf A21}, L459 (1988).

\bibitem{bound}
L. Kirousis, E. Kranakis, D. Krizanc, in
{\em Proceedings of the 4th European Symposium on Algorithms},
27-38 (1992). \\
O. Dubois, Y. Boufkhad, {\em Journal of Algorithms} 
{\bf 24}, 395--420 (1997).

\bibitem{remark:frozen} The ground state entropy is finite due to the
existence of under-contrained spins, {\it i.e.} within our approach it
is bounded at $\alpha_c$ by $(1-B_0-B_1) \ln 2\approx 0.055$ from
above. This value coincides within our numerical precision with the
SAT entropy $s_{rsb}(\alpha=4.48)\approx 0.054$, but is smaller than
the entropy $s_{rsb} \approx 0.073 - 0.081$ at the numerical estimates
for $\alpha_c=4.25-4.3$.  The latter entropy values would require at
least $10-12\%$ of under-constrained spins.
 
\bibitem{zero.entropy} In fact, a zero-entropy criterion is
frequently used in discrete infinite connectivity models in order to 
locate $T=0$ phase transitions, {\it e.g.} in the case of the learning
behavior of binary neural networks \cite{KrMe}.

\bibitem{zagrebnov}
U. Behn, J.L. van Hemmen, R. K{\"u}hn, A. Lange, V.A. Zagrebnov, {\it
Physica D} {\bf 68}, 401 (1993).

\bibitem{KrMe} W. Krauth and M. M{\'e}zard, {\it J. Phys. (France)}
{\bf 50}, 3059 (1989).\\
G. Gy{\"o}rgyi, {\it Phys. Rev. Lett.} {\bf 64}, 2957 (1990).

\end{thebibliography}
\end{document}